\documentclass[11pt,a4paper]{article}

\usepackage{jheppub}
\usepackage{mathtools,bm,amssymb,mathrsfs}
\usepackage{bookmark}
\usepackage{tikz}
\usetikzlibrary{arrows.meta,decorations.markings}

\hypersetup{
  pdftitle={Perturbative Anomaly Inflow on Orbifolds},
  pdfauthor={Hao Y. Zhang},
  pdfsubject={E8 orbi-instantons, Donnelly equivariant APS theorem, anomaly inflow}
}

\allowdisplaybreaks

\usepackage{xcolor}

\newcommand{\ch}{\operatorname{ch}}
\newcommand{\Tr}{\operatorname{Tr}}
\newcommand{\tr}{\operatorname{tr}}
\newcommand{\STr}{\operatorname{STr}}

\newcommand{\Ad}{\operatorname{Ad}}
\newcommand{\rank}{\operatorname{rank}}
\newcommand{\ind}{\operatorname{ind}}

\newcommand{\dd}{\mathrm d}
\newcommand{\Z}{\mathbb Z}
\newcommand{\Rset}{\mathbb R}
\newcommand{\C}{\mathbb C}

\newcommand{\one}{\mathbf 1}

\IfFileExists{references.bib}{}{%
  \AtBeginDocument{\renewcommand{\cite}[1]{[\nolinkurl{##1}]}}%
}

\title{\boldmath Perturbative Anomaly Inflow on Orbifolds}
\author{Hao Y. Zhang}

\affiliation{
    Kavli Institute for the Physics and Mathematics of the Universe (WPI), \\
    University of Tokyo, Kashiwa, Chiba 277-8583, Japan \\}

\emailAdd{hao.zhang@ipmu.jp}

\abstract{We study perturbative anomaly inflow for effective theories
supported on orbifold fixed loci of the internal geometry. The equivariant APS theorem supplies a Donnelly fixed-point density. We identify its degree-\((d+2)\) component as the anomaly polynomial of the $d$-dimensional effective theory.
We apply the construction to six-dimensional A-type orbi-instanton
theories engineered by \(N\) M5-branes probing a transverse
\(\C^2/\Z_k\) singularity at an end-of-the-world M9-brane. For a flat
\(E_8\) connection specified by \(\rho:\Z_k\to E_8\), we compute the
non-identity ALE fixed-point class on the M9 wall. Its degree-eight
component agrees with the complete Kac-label-dependent remainder
conjectured in \cite{MOTZ}, including the \(SU(2)_R\) and tangent-bundle
curvatures. Together with the known M5/Ho\v{r}ava--Witten and
ALE/Ho\v{r}ava--Witten terms, this reproduces the unflavored
tensor-branch anomaly polynomial. This local equality further suggests a complete M-theory corner-inflow interpretation. The same prescription also be extended to include flavor symmetries as the centralizer of $E_8$.}

\keywords{anomaly inflow, APS \(\eta\)-invariant, six-dimensional SCFTs, orbi-instantons}

\begin{document}
\maketitle

\section{Introduction}
\label{sec:introduction}

\subsection{Anomaly inflow and index theory}
\label{sec:intro-index-history}

\paragraph{Perturbative anomalies and inflow.}
For a chiral fermion in even dimensions, the perturbative anomaly is encoded by an anomaly polynomial \(I_{d+2}\). Locally, it obeys the descent equations
\begin{equation}
 I_{d+2}=\dd I_{d+1}^{(0)},
 \qquad
 \delta I_{d+1}^{(0)}=\dd I_d^{(1)}.
\end{equation}
Here \(I_{d+1}^{(0)}\) is a Chern--Simons form in one higher dimension, while \(I_d^{(1)}\) gives the gauge or local Lorentz variation of the effective action \cite{AGW,GreenSchwarz}. The boundary anomaly can therefore be cancelled by the boundary variation of a bulk term \(\int I_{d+1}^{(0)}\). Such a descent formalism is a core mechanism of anomaly inflow.

Callan and Harvey gave a direct spacetime realization of this mechanism in their analysis of effective strings and domain-wall fermions \cite{CallanHarvey}. When a fermion mass changes sign across a defect, the Dirac equation develops chiral zero modes localized on the defect. Taken by themselves, these modes are anomalous. However, integrating out the massive bulk fermion induces Chern--Simons terms with different coefficients on the two sides. Their variation transports precisely the current needed to cancel the zero-mode anomaly. The chiral defect anomaly and the parity-odd bulk Chern--Simons term are thus the boundary and bulk descriptions of the same fermionic system at low energies.

\paragraph{The \texorpdfstring{\(\eta\)}{eta}-invariant.}
Chern--Simons forms capture only the local, perturbative part of an
anomaly. The Dai--Freed formalism gives a more general description.
A complex chiral fermion defines a section of a determinant line,
whereas a chiral fermion with an appropriate reality condition
defines a section of a Pfaffian line. Their anomalies are the
obstructions to trivializing the corresponding line bundles
\cite{DaiFreed,FreedDeterminants}. Define the reduced eta-invariant by
\[
 \xi(D_Y)=\frac{\eta(D_Y)+h(D_Y)}2,
 \qquad h(D_Y):=\dim\ker D_Y .
\]
With the phase conventions used below, the complex and real bulk
phases are respectively
\begin{equation}
 Z_{\rm bulk}^{\mathbb C}(Y)=\exp[-2\pi i\,\xi(D_Y)],
 \qquad
 Z_{\rm bulk}^{\mathbb R}(Y)=\exp[-\pi i\,\xi(D_Y)] .
 \label{eq:intro-complex-real-eta-phases}
\end{equation}
The second phase is the Pfaffian square root of the first whenever the
relevant real structure defines a Pfaffian theory. The APS index
theorem relates these spectral phases to a local index density. It
thereby explains why their continuous variation is governed by a
Chern--Simons term, while their discrete information can detect
global anomalies \cite{APS1,APS2,YonekuraDaiFreed,WittenYonekura}.

\paragraph{Orbifolds.}
We consider compactifications in which the internal space is a quotient by a finite group $G$. The fixed loci in the total space host the lower-dimensional effective theory, so we need to upgrade our formalism so that the eta-invariant w.r.t. the internal directions also carries information about the orbifold projection. This is achieved by inserting the group averaging
\[
 P_G=\frac{1}{|G|}\sum_{g\in G}g
\]
into the trace over the spectrum of the Dirac operator, giving an \textit{equivariant} eta-invariant. In the bulk, the index density of the chiral Dirac operator localizes to the fixed set of each \(g\), as can be made concrete using equivariant \(K\)-theory. Donnelly's equivariant APS theorem further computes the equivariant index in terms of the resulting bulk density combined with the boundary spectral term \cite{AtiyahSegalII,AtiyahBott,Donnelly}. Focusing on the perturbative part for now, the integration over the internal orbifold directions (mathematically, the pushforward along the fixed fibers) then produces the anomaly polynomial on the external spacetime. The construction may be summarized as
\[
 \begin{gathered}
 \text{projection by the orbifold group}
 \\[-2pt]
 \Big\downarrow
 \\
 \text{localization of the Dirac symbol}
 \\[-2pt]
 \text{via equivariant \(K\)-theory}
 \\[-2pt]
 \Big\downarrow
 \\
 \text{integration over the orbifold directions}.
 \end{gathered}
\]
Section~\ref{sec:general-analysis} gives the geometric setup and the precise form of the equivariant APS theorem used below.

\subsection{Application: anomalies of A-type orbi-instanton theories}
\label{sec:intro-application}

As a stringent test, we derive a uniform non-identity ALE fixed-point
contribution on the M9 wall and show that it agrees with the
Kac-label-dependent remainder conjectured in \cite{MOTZ}.
We regard the fixed-point construction as a physical prescription
rather than a mathematical derivation. We expect a complete proof for
the wall problem to require the equivariant family index theorem, but
we omit that proof and instead test the prescription through the
anomaly-polynomial matching carried out below.

\paragraph{Boundary data in M-theory and 6D SCFTs.}
Six-dimensional superconformal field theories (6D SCFTs) are strongly coupled theories which often lack a conventional Lagrangian description; see \cite{HeckmanRudeliusReview} for a review. Their geometric building blocks and gluing rules were organized in the atomic classification of Heckman, Morrison, Rudelius, and Vafa \cite{Atomic6D}. A broad class of 6D SCFTs admits an M-theory realization with additional algebraic data specifying "boundary conditions". At an ADE singularity one may turn on a ``T-brane" vacuum expectation value, described by a nilpotent homomorphism \(\mu:\mathfrak{su}(2)\to G_{\mathrm{ADE}}\)\cite{Cecotti:2010bp}, while the M9 worldvolume may carry a flat \(E_8\) connection specified by \(\rho:\Gamma_{\mathrm{ADE}}\to E_8\) \cite{HMVClassification}. The relation between these algebraic data and tensor-branch descriptions has been studied systematically \cite{NilpotentHierarchies,FissionFusion,TBraneJunctions,MRTTBranes,MOTZ}. 

\begin{figure}[htbp]
 \centering
 \includegraphics[width=\textwidth]{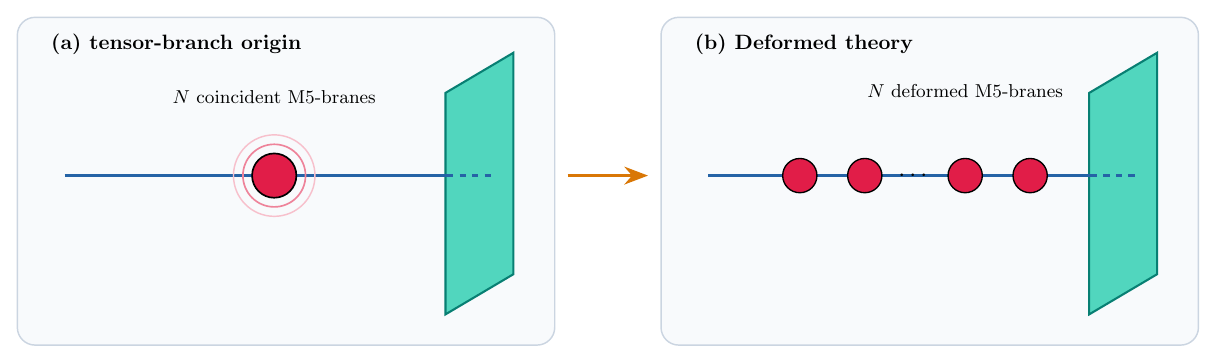}
 \caption{M-theory construction of an A-type orbi-instanton theory. The blue line depicts the singular locus of the \(\C^2/\Gamma_{A_{k-1}}\cong\C^2/\Z_k\) ; its dashed segment ends on the green end-of-the-world M9-brane. Red dots denote M5-branes and the orange arrow denotes the tensor-branch deformation. (a) At the origin of the tensor branch, the \(N\) M5-branes coincide on the singularity. (b) A tensor-multiplet expectation value separates the M5-branes in the transverse direction, while the ALE singularity and the M9-brane remain fixed.}
 \label{fig:m-theory-tensor-branch}
\end{figure}

\paragraph{Orbi-instanton theories.}
Figure~\ref{fig:m-theory-tensor-branch} shows the M-theory configuration used throughout this paper. At the conformal point, the M5-branes coincide at the intersection of the M9-brane and the \(A_{k-1}\) singularity. A tensor-branch expectation value separates the M5-branes in the transverse direction. We set \(\mu\) to the trivial nilpotent orbit and only consider a general flat \(E_8\) connection specified by \(\rho\). For \(\Gamma_A=\Z_k\), $\rho$ is classified combinatorically by Kac label, and the tensor branch is known completely and falls into five infinite families according to \(N\), \(k\), and \(\rho\) \cite{MOTZ}. 

\paragraph{Conjectural form of anomaly polynomial.}
It is generally difficult to extract observables from a six-dimensional SCFT, but its anomaly polynomial is one of the robust quantities. Moving onto the tensor branch by giving Vacuum Expectation Values (VEVs) to tensor-multiplet scalars yields a low-energy description in which the fermions inside hypermultiplets, vector multiplets, and tensor multiplets all contribute one-loop anomalies. The six-dimensional Green--Schwarz--Sagnotti--West (GSSW) mechanism \cite{GSSW,SagnottiGS} then cancels the anomalies of the dynamical gauge fields, leaving the eight-form anomaly of the ultraviolet SCFT \cite{OSTeString,OSTY6d,Intriligator6d}. For A-type orbi-instanton theories, tensor-branch and compactification computations give a detailed homomorphism-dependent result \(I_8^{N,k}(\rho)\) \cite{MOTZ}, but no direct M-theory inflow derivation had been available. The relevant even-dimensional fermions live on the ten-dimensional M9 worldvolume, within which the asymptotic boundary of the internal space can be viewed as a three-dimensional Lens space. Reference~\cite{MOTZ} identified two established inflow terms inside \(I_8^{N,k}(\rho)\): one from the system of \(N\) M5-branes ending on the M9-brane, namely the E-string system, and one from \(N\) M5-branes probing the ADE singularity, namely the conformal-matter system \cite{ConformalMatter}. Denoting their sum by \(I_8^{\mathrm{naive\ inflow}}\), one may write
\[
 I_8^{N,k}(\rho)=I_8^{\mathrm{naive\ inflow}}+c(\rho).
\]
We show that the non-identity ALE fixed-point contribution on the M9
wall agrees exactly with the polynomial \(c(\rho)\) inferred in
\cite{MOTZ}. This fixed-point class is supported at the intersection
of the wall and the ALE fixed plane and retains the full \(E_8\)
character of \(\rho\). This equality supports the proposed
interpretation of \(c(\rho)\) as a codimension-five corner inflow
term. Establishing equality of the complete M-theory anomaly lines
would additionally require a bulk--wall--ALE gluing argument
including the \(C\)-field and the corner boundary conditions.

\subsubsection{The conjectural formula for \texorpdfstring{\(I_8(\rho)\)}{I8(rho)}}
\label{sec:intro-i8-summary}

For \(N\) M5-branes at an \(A_{k-1}\) singularity ending on the
\(E_8\) wall, the homomorphism \(\rho:\Z_k\to E_8\) is encoded by
Kac data \(w\). We use \(N:=N_3\), where \(N_3\) is the rank
parameter of \cite{MOTZ}; adding one M5-brane increases this parameter
by one. Other charge variables used in that reference can differ from
\(N\) by convention-dependent shifts. Tensor-branch computations
lead to the unified decomposition
\begin{equation}
 I_8^{\rm TB}(\rho)=I_8^{\rm inflow,naive}\!\left(Q(\rho)\right)+c(\rho),
 \qquad
 Q=N+\frac12\left(k+\frac1k-\frac{\langle w,w\rangle}{k}\right).
 \label{eq:intro-i8-summary}
\end{equation}
Here \(I_8^{\rm inflow,naive}\!\left(Q(\rho)\right)\) is the sum of the E-string and conformal-matter inflow terms. Choose a maximal torus of \(E_8\) and write
\begin{equation}
 h_\rho=\rho(1)=\exp\!\left(\frac{2\pi i}{k}w\right),
 \qquad
 w=\sum_{i=1}^{8}n_i\omega_i^\vee.
 \label{eq:intro-kac-holonomy}
\end{equation}
Let \(\Delta_+\) be the set of positive roots and define the \(E_8\) Weyl vector by
\begin{equation}
 \rho_{\rm W}:=\frac12\sum_{\alpha\in\Delta_+}\alpha.
 \label{eq:intro-Weyl-vector}
\end{equation}
We also set \(p_i=p_i(TW_6)\) and \(c_2(R)=\Tr F_R^2\). The complete Kac-polynomial remainder inferred in \cite{MOTZ} is
\begin{equation}
 c(\rho)=\frac1k\bigl(P_0+P_2+P_4+P_6\bigr)
 +\frac12 I_{\rm vector}^{\rm free},
 \label{eq:intro-MOTZ-inductive}
\end{equation}
where
\begin{align}
 P_0={}&\frac1{384}
 \bigl(-88c_2(R)^2+32c_2(R) p_1-5p_1^2+4p_2\bigr),
 \label{eq:intro-MOTZ-P0}\\
 P_2={}&\frac{k^2}{11520}
 \bigl(2512c_2(R)^2-760c_2(R) p_1+157p_1^2-124p_2\bigr)\notag\\
 &+\frac{15\langle w,w\rangle-k\langle w,\rho_{\rm W}\rangle}{5760}
 \bigl(112c_2(R)^2-40c_2(R) p_1+7p_1^2-4p_2\bigr),
 \label{eq:intro-MOTZ-P2}\\
 P_4={}&-\frac1{288}
 \left(
 9\langle w,w\rangle^2+15k^2\langle w,w\rangle-2k^4
 -k\sum_{\alpha\in\Delta_+}\langle w,\alpha\rangle^3
 \right)\notag\\[-1mm]
 &\hspace{29mm}\times\bigl(4c_2(R)^2-c_2(R) p_1\bigr),
 \label{eq:intro-MOTZ-P4}\\
 P_6={}&\frac1{240}
 \left(
 5\langle w,w\rangle^3+15k^2\langle w,w\rangle^2
 -5k^4\langle w,w\rangle+k^6
 -k\sum_{\alpha\in\Delta_+}\langle w,\alpha\rangle^5
 \right)c_2(R)^2,
 \label{eq:intro-MOTZ-P6}\\
 I_{\rm vector}^{\rm free}={}&\frac1{5760}
 \bigl(-240c_2(R)^2-120c_2(R) p_1-7p_1^2+4p_2\bigr).
 \label{eq:intro-MOTZ-free-vector}
\end{align}
If \(w\) is assigned degree one in \(k\), then \(P_{2m}\) is
homogeneous of degree \(2m\) in the Kac data.
Section~\ref{sec:tensor-review} rewrites the full formula before
comparing it with the ALE fixed-point computation on the M9 wall. The central
question is whether the complete \(\rho\)-dependent remainder can be
reproduced by a single equivariant fixed-point sum without working
through the tensor-branch quivers case by case.
\par\noindent\mbox{}\par\vspace{-\baselineskip}

We remark that the tensor branch falls into five general families
depending on \(N\), \(k\), and the Kac label. The existence of a
uniform polynomial is therefore highly non-trivial. Moreover, the
formula in \cite{MOTZ} contains high-degree power sums over the
positive roots of \(E_8\). These expressions provide both the
guidance and a stringent algebraic check for the uniform M9
fixed-point computation developed below.

\subsubsection{Outline of the inflow derivation}

\paragraph{Established inflow terms.}
As reviewed in \cite{MOTZ}, the known classical inflow for A-type orbi-instanton theories splits according to its physical origin:
\begin{equation}
 I_8^{\rm naive}=I_8^{\rm M5/HW}+I_8^{\rm ALE/HW}.
 \label{eq:intro-naive-split}
\end{equation}
The first term follows from the eleven-dimensional Chern--Simons coupling, the M5-brane magnetic source, and the Ho\v{r}ava--Witten boundary condition. The second comes from the seven-dimensional ALE fixed plane and its boundary condition at the wall. Lifting the flux from the quotient to the covering space replaces the charge by \(kQ\), while the integral acquires a factor \(1/k\). Hence
\begin{equation}
 I_8^{\rm M5/HW}(Q)=\frac1k I_{8,\rm cover}^{\rm M5/HW}(kQ).
 \label{eq:intro-cover-identity}
\end{equation}
This identity accounts for the terms in the naive inflow that are cubic, quadratic, and linear in the M5-brane charge \(Q\). Their explicit form is given in the main text.

\paragraph{The additional fixed-point term.}
The orbifold-projected wall one-loop class decomposes into its
identity and non-identity sectors:
\begin{equation}
 I_{\rm wall}^{\rm proj}(\rho)
 =\frac1k I_{\rm wall}^{\rm cover}+c_{\rm wall}(\rho).
 \label{eq:intro-wall-sector-decomposition}
\end{equation}
Here
\begin{equation}
 c_{\rm wall}(\rho)
 :=\left[\widehat A(TW)\frac1k\sum_{j=1}^{k-1}
 K_j(r)\ch_{g^j,h_\rho^j}(V_{\rm HW})\right]_8.
 \label{eq:intro-wall-sum}
\end{equation}
Here \(K_j(r)\) is the Atiyah--Bott normal factor obtained as the ratio
of the normal-spinor virtual class to the self-intersection Euler
class, while \(h_\rho=\rho(g)\) acts on the wall \(E_8\) gaugino
bundle.

We assume that the usual smooth Ho\v{r}ava--Witten cancellation
commutes with passage to the quotient, so that the identity sector
obeys
\begin{equation}
 \frac1k\left(I_{\rm wall}^{\rm cover}
 +I_{\rm HW,bulk}^{\rm cover}\right)=0.
 \label{eq:intro-identity-sector-cancellation}
\end{equation}
The computation below determines the remaining class
\(c_{\rm wall}(\rho)\). For notational economy, after establishing
its equality with the polynomial of \cite{MOTZ}, we also denote it by
\(c(\rho)\).

The two terms in the candidate anomaly polynomial have distinct
origins. The naive term contains the established
M5/Ho\v{r}ava--Witten and ALE/Ho\v{r}ava--Witten contributions, while
\(c_{\rm wall}\) is the non-identity ALE fixed-point class on the M9 wall. Ignoring
the unbroken part of the internal \(\mathfrak{su}(2)_L\), define
\begin{equation}
 I_8^{\rm cand}(\rho;F_H,F_R,F_{SU(k)})
 :=I_8^{\rm naive}\!\left(Q(\rho);F_H,F_R,F_{SU(k)}\right)
 +c_{\rm wall}(\rho;F_H).
 \label{eq:intro-main-result}
\end{equation}
We show that this candidate equals the perturbative tensor-branch
anomaly polynomial. This proves an equality of local eight-form
classes, while the corresponding global anomaly analysis
is left for future work.

Section~\ref{sec:general-analysis} reviews anomaly inflow and eta
invariants and develops the equivariant fixed-point framework for an
internal orbifold with asymptotic boundary.
Section~\ref{sec:inflow} reviews the conjectural tensor-branch
polynomial, derives the non-identity ALE fixed-point class on the M9 wall, compares
it term by term with the Kac polynomial, and incorporates continuous
backgrounds for the wall centralizer inside \(E_8\).
Section~\ref{sec:discussion} discusses the precise scope of the
perturbative result and its relation to the complete M-theory quantum
phase. Appendix~\ref{app:aps-to-donnelly} derives the fixed-point
normal factor from the equivariant Dirac symbol. 

\section{General analysis of anomaly inflow and orbifold backgrounds}
\label{sec:general-analysis}

We begin by reviewing the Dai--Freed anomaly theory and anomaly inflow on a smooth background, and then extend it to the case when the internal manifold takes the form of an orbifold.

\subsection{Fermionic anomaly inflow and the \texorpdfstring{\(\eta\)}{eta}-invariant}
\label{sec:dai-freed}

\subsubsection{Perturbative anomaly inflow}
\label{sec:perturbative-inflow-review}

The local gauge and gravitational anomalies of an even-dimensional chiral fermion are encoded by an anomaly polynomial \(I_{d+2}\). After choosing a local Chern--Simons form \(I_{d+1}^{(0)}\), the descent equations read
\begin{equation}
 I_{d+2}=\dd I_{d+1}^{(0)},
 \qquad
 \delta I_{d+1}^{(0)}=\dd I_d^{(1)}.
 \label{eq:perturbative-descent}
\end{equation}
If the \(d\)-dimensional theory lives on \(Y_d=\partial X_{d+1}\), the anomalous variation of its effective action is
\begin{equation}
 \delta\Gamma_Y
 =2\pi\mathrm i\int_Y I_d^{(1)}.
 \label{eq:boundary-perturbative-anomaly}
\end{equation}
Adding the bulk term
\begin{equation}
 S_{\mathrm{CS}}[X]
 =-2\pi\mathrm i\int_X I_{d+1}^{(0)}
 \label{eq:bulk-CS-inflow}
\end{equation}
gives \(\delta S_{\mathrm{CS}}=-\delta\Gamma_Y\) by Stokes' theorem. The boundary effective action need not be gauge invariant on its own; gauge invariance is required only of the product of the boundary fermion partition function and the bulk Chern--Simons phase. In this way the anomaly is naturally realized as inflow from one dimension higher \cite{AGW,GreenSchwarz}.

The Callan--Harvey mechanism provides the microscopic fermionic picture \cite{CallanHarvey}. Consider a massive Dirac fermion in an odd-dimensional spacetime with transverse coordinate \(s\), and let its mass \(m(s)\) change sign across a domain wall localized at $s = 0$. Writing \(D_Y\) for the Dirac operator along the wall, the equation is
\begin{equation}
 \left(D_Y+\gamma^\perp\partial_s-m(s)\right)\psi=0.
 \label{eq:CH-domain-wall-dirac}
\end{equation}
If \(m(s)\) changes sign from negative to positive near \(s=0\), there is a normalizable zero mode localized on the wall,
\begin{equation}
 \psi(x,s)
 =\chi(x)
 \exp\!\left[-\int_0^s m(s')\,\dd s'\right],
 \qquad
 D_Y\chi=0.
 \label{eq:CH-wall-zero-mode}
\end{equation}
The transverse Dirac equation retains only one chirality of \(\chi\), determined by the direction of the mass jump. A nonchiral bulk fermion therefore produces an even-dimensional chiral fermion on the wall, and the zero mode living on the wall is anomalous on its own. Integrating out the massive modes on either side of the wall then produces a parity-odd Chern--Simons effective action, whose coefficient depends on \(\operatorname{sign}m(s)\) and therefore jumps at the wall. If \(k(s)\) denotes this piecewise constant coefficient, then
\begin{equation}
 \Gamma_{\mathrm{odd}}
 =2\pi\mathrm i\int_X k(s)I_{d+1}^{(0)},
 \qquad
 \delta\Gamma_{\mathrm{odd}}
 =-2\pi\mathrm i\int_X
 \dd k\wedge I_d^{(1)}.
 \label{eq:CH-CS-jump}
\end{equation}
Because \(\dd k\) is supported at \(s=0\), the variation of the bulk term localizes on the wall and cancels the anomaly of the chiral zero mode. Equivalently, the massive bulk modes generate a transverse current flowing into the wall whose divergence compensates for the nonconservation of the wall current.

The same mechanism appears on the codimension-two defect in the Callan--Harvey effective-string model. The phase of a complex fermion mass winds around the string, so zero modes of the transverse Dirac operator become chiral fermions on the string worldsheet. After the massive modes away from the string get integrated out, a bulk coupling between the mass phase and the gauge field produces radial inflow whose localized part cancels the two-dimensional zero-mode anomaly. The common lesson of the domain-wall and effective-string examples is that the chiral defect modes and the parity-odd effective action of the massive bulk fermion arise from a single Dirac operator and, therefore, cannot be regulated independently---their regulators need to be compatible with one another so as not to generate an arbitrary phase factor.

We use the term \emph{perturbative anomaly} for the local curvature of the anomaly line bundle, namely the part described by \(I_{d+2}\) and descent. A \emph{global} or \emph{nonperturbative anomaly} is the holonomy of a flat anomaly line around a noncontractible loop in parameter space. Here ``nonperturbative'' means that the phase cannot be detected by an infinitesimal local variation and is unrelated to the coupling strength.

\subsubsection{Global anomalies and the \texorpdfstring{\(\eta\)}{eta}-invariant}
\label{sec:nonperturbative-eta-review}

A local Chern--Simons action reproduces the perturbative anomaly but
does not, in general, determine the full phase of the fermion path
integral. A complex fermion produces a determinant, while a fermion
with an appropriate reality condition produces a Pfaffian, with
\((\operatorname{Pf}D)^2=\det D\). For a chiral fermion, the
determinant or Pfaffian is generally a section of a line bundle rather
than a globally defined complex-valued function.

Let
\begin{equation}
 Y_d=\partial X_{d+1}.
 \label{eq:Y-boundary-X}
\end{equation}
With the conventions of
equation~\eqref{eq:intro-complex-real-eta-phases}, the regulated
bulk--boundary partition functions are
\begin{align}
 Z_{\mathbb R}(Y;X)
 &=\left|\operatorname{Pf}D_Y^+\right|
 \exp[-\pi i\,\xi(D_X)]\notag\\
 &=\left|\operatorname{Pf}D_Y^+\right|
 \exp\!\left[-\frac{\pi i}{2}
 \bigl(\eta(D_X)+h(D_X)\bigr)\right],
 \label{eq:WY-inflow-real}\\
 Z_{\mathbb C}(Y;X)
 &=\left|\det D_Y^+\right|
 \exp[-2\pi i\,\xi(D_X)].
 \label{eq:WY-inflow-complex}
\end{align}
On a patch where the bulk operator is invertible, \(h(D_X)=0\), and
the real phase reduces to \(\exp[-\pi i\eta(D_X)/2]\).

Neither expression should be regarded as a product of two
independently defined complex numbers. The boundary determinant or
Pfaffian is a section of an anomaly line \(\mathcal L\), while the
properly defined Dai--Freed bulk phase is a section of
\(\mathcal L^{-1}\). Only their product is a well-defined number.
Changing the filling can change the resulting invertible bulk phase
and therefore contains global information not captured by a local
anomaly polynomial
\cite{DaiFreed,WittenYonekura,YonekuraDaiFreed}.

\paragraph{Recovering the perturbative anomaly.}
For infinitesimal variations, the Dai--Freed description reduces to
the usual descent equations. The curvature of the Bismut--Freed
connection on the determinant or Pfaffian line satisfies
\begin{equation}
 \frac{F_{\mathcal L}}{2\pi\mathrm i}
 =
 \left[\int_{Y_d}I_{d+2}\right]_{(2)}.
 \label{eq:BF-curvature}
\end{equation}
Here \(I_{d+2}\) is understood in the same fermion convention as the
anomaly line: the anomaly polynomial of a real/Pfaffian fermion is one
half of the corresponding complex-Weyl index density. The expression
on the right is first integrated over \(Y_d\), after which its
two-form component on parameter space is extracted. Thus
\(I_{d+2}\) remains the spacetime anomaly polynomial, while
equation~\eqref{eq:BF-curvature} identifies it with the local
curvature of the anomaly line
\cite{BismutFreed,FreedDeterminants}. Its infinitesimal holonomy is
equivalently described by the descent of the corresponding
Chern--Simons functional.

\paragraph{Global anomalies and the eta-invariant.}
The eta-invariant in the phase above follows directly from
Pauli--Villars regularization. In the real-fermion convention,
integrating out a nonzero bulk eigenmode \(\lambda\) contributes a
phase proportional to \(-\pi\operatorname{sign}(\lambda)/2\); the
corresponding complex-fermion phase is twice this contribution.
Formally summing over all modes gives the difference between the
number of positive and negative eigenvalues. If \(M\) is a closed
odd-dimensional spin manifold and \(D_M\) is a self-adjoint Dirac
operator, one would formally write
\begin{equation}
 \eta_{\mathrm{formal}}(D_M)
 =
 \#\{\lambda>0\}-\#\{\lambda<0\}
 =
 \sum_{\lambda\ne0}\operatorname{sign}(\lambda),
 \label{eq:eta-sign-sum-formal}
\end{equation}
which gives the $\eta$-invariant by definition. But since we usually get infinitely many positive and negative eigenvalues, this expression must be regularized. For \(\operatorname{Re}(s)\) sufficiently large, we define
\begin{equation}
 \eta(D_M,s)
 =
 \sum_{\lambda\ne0}
 \operatorname{sign}(\lambda)|\lambda|^{-s},
 \qquad
 \eta(D_M):=\eta(D_M,0),
 \label{eq:eta-zeta-main}
\end{equation}
where the value at \(s=0\) is obtained by analytic continuation.

The Mellin transform relates spectral asymmetry to local geometry:
\begin{equation}
 \eta(D_M,s)
 =
 \frac{1}{\Gamma((s+1)/2)}
 \int_0^\infty
 t^{(s-1)/2}
 \Tr\!\left(D_Me^{-tD_M^2}\right)\dd t.
 \label{eq:eta-heat-main}
\end{equation}
Large \(t\) probes the low-lying spectrum, while the small-\(t\) region is controlled by high-energy modes and admits a local heat-kernel asymptotic expansion. This representation contains both the full Dirac spectrum and its short-distance local expansion, thereby connecting the global spectral phase with the perturbative anomaly polynomial.

Let
\[
 h(D_M)=\dim\ker D_M.
\]
The reduced APS invariant is
\begin{equation}
 \xi(D_M)
 =
 \frac{\eta(D_M)+h(D_M)}{2}.
 \label{eq:xi-def}
\end{equation}
When an eigenvalue crosses zero, \(\eta\) and \(h\) jump separately, but \(\xi(D_M)\bmod\mathbb Z\) remains continuous. The correction to the anomaly phase is given by the \textit{reduced} eta-invariant according to the Atiyah--Patodi--Singer theorem, as we now explain.

\paragraph{The APS index theorem.}
Introduce an auxiliary even-dimensional manifold \(Z\) with \(M=\partial Z\). When \(M\) is the asymptotic boundary \(Y=\partial X\) of the internal space, its filling \(Z\) is arbitrary and need not coincide with \(X\); we thus use a separate set of notation to emphasize such a distinction coming from the arbitrariness of $Z$. The chiral Dirac operator with APS boundary conditions satisfies
\begin{equation}
 \operatorname{ind}(D_Z^+)
 =
 \int_Z\widehat A(TZ)\operatorname{ch}(E)
 -\xi(D_M).
 \label{eq:APS-main}
\end{equation}
Here \(D_Z^+\) maps positive-chirality spinors to negative-chirality spinors, and its index is defined as
\[
 \operatorname{ind}(D_Z^+)
 =
 \dim\ker D_Z^+
 -\dim\operatorname{coker}D_Z^+.
\]
Because the index is integral, its exponential does not contribute to the continuously varying anomaly phase. As far as anomalies are concerned, equation~\eqref{eq:APS-main} therefore relates the boundary spectral invariant to the bulk index density. Locally, it reproduces the anomaly polynomial and its Chern--Simons descent; globally, the eta-invariant retains spectral information that cannot be recovered from local differential forms.

A large gauge transformation or large diffeomorphism defines a noncontractible loop in parameter space. Gluing the initial and final physical spacetimes by that transformation produces a closed mapping torus. The holonomy of the anomaly line around the loop is the eta-phase on the mapping torus. This phase can remain nontrivial even when the line-bundle curvature vanishes and therefore detects global anomalies invisible to the perturbative anomaly polynomial. The computations in our paper only use the perturbative part, namely the eight-form \(I_8\) associated with the 6D spacetime. We now introduce an orbifold quotient in the internal space and determine how to describe the anomaly inflow in such a scenario with a modified version of the APS index theorem.

\subsection{Equivariant APS index theorem and inflow on orbifold backgrounds}
\label{sec:orbifold-aps}

We now extend our anomaly inflow analysis to compactifications with an internal
orbifold singularity. The quotient by the finite group $G$ changes both the boundary spectrum and the
bulk index density: only $G$-invariant modes survive on the entire internal space, while the non-identity
terms in the projector localize the index density on the fixed set in
equivariant \(K\)-theory, with an additional factor associated with such localization. 

We take the external Euclidean spacetime
\(M_d\) to be closed. On the covering space,
\begin{equation}
 X=M_d\times Z_{2r},\qquad
 Y=\partial X=M_d\times L_{2r-1},\qquad
 L_{2r-1}=\partial Z_{2r}.
 \label{eq:general-total-space}
\end{equation}
If \(M_d\) had a boundary, \(X\) would also have the boundary face
\(\partial M_d\times Z_{2r}\) and the corner
\(\partial M_d\times L_{2r-1}\); an index theorem with corners would then
be required, which we do not consider in our paper.

Near a thin collar-shaped neighborhood of the boundary, the chiral Dirac operator on the
even-dimensional \(X\) is
\[
 D_X=\gamma^\perp(\partial_u+D_Y),
\]
where \(D_Y\) is the self-adjoint Dirac operator on the
odd-dimensional \(Y\); its spectral asymmetry gives precisely the APS
\(\eta\)-correction.
The finite group \(G\) acts only in the internal directions and its action
is then lifted to the spinor and twisted gauge bundles. We assume that non-identity
elements act freely on \(L_{2r-1}\), although they may have fixed points in
\(Z_{2r}\). If a non-identity element also fixes points on
\(L_{2r-1}\), then \(Y/G\) is itself a stratified orbifold boundary, and
the degrees of freedom and boundary data on its singular strata must be
included. Orbifold singularities extending to the asymptotic boundary and
their cutting-and-gluing structure are analyzed in
\cite{CveticCuttingGluing}; the extra-dimensional \(\eta\)-invariant
description of non-isolated singularities is developed in
\cite{CveticExtraEta}, and a generalization to quiver gauge theories was carried out in \cite{Chakrabhavi:2026iku}, focusing on anomalies of higher-form symmetries. The Donnelly formula used here is restricted to an orbifold action which acts freely on the boundary.

\subsubsection{From orbifold projection to the fixed-point anomaly}
\label{sec:orbifold-inflow-summary}

Before introducing the group action, recall the ordinary non-equivariant starting point. An elliptic operator \(P\) on a closed compact manifold \(X\) has a principal symbol that is invertible away from the zero section and therefore defines a compactly supported class \([\sigma(P)]\in K_c^0(T^*X)\). The Atiyah--Singer theorem identifies its analytic index with the \(K\)-theory pushforward of this class:
\begin{equation}
 \operatorname{ind}_{\rm an}(P)
 =\operatorname{ind}_{\rm top}\!\left([\sigma(P)]\right)
 =p_![\sigma(P)]
 \in K^0(\mathrm{pt})\cong\mathbb Z,
 \qquad p:T^*X\longrightarrow\mathrm{pt}.
 \label{eq:ordinary-symbol-pushforward-main}
\end{equation}
For a twisted chiral Dirac operator this pushforward becomes
\begin{equation}
 \operatorname{ind}(D_{X,E}^+)
 =\int_X\widehat A(TX)\operatorname{ch}(E).
 \label{eq:ordinary-dirac-index-main}
\end{equation}
Thus the symbol is not introduced only for the orbifold problem: it is already the \(K\)-theoretic datum whose pushforward gives the ordinary index. If \(X\) has a boundary, the symbol still fixes the local bulk density, but the complete Fredholm problem also requires a boundary condition and the APS spectral correction. The equivariant construction replaces \(K_c^0(T^*X)\) by \(K_{G,c}^0(T^*X)\) and the integer-valued index by a virtual \(G\)-representation; localization at a non-identity element then restricts the pushforward to its fixed set. Appendix~\ref{app:aps-to-donnelly} gives the detailed derivation using the symbol class, the self-intersection formula, and the fixed-point normal factor.

\begin{itemize}
 \item \textbf{One-loop behavior and projection of the boundary spectrum.}
 Integrating out a quadratic fermion produces a Pfaffian or determinant. Its
 dependence on the background connection and metric is what we call its
 one-loop behavior. Since the orbifold quotient retains only \(G\)-invariant
 modes, the projector must be inserted into the boundary spectral trace:
 \begin{equation}
  \Pi_G=\frac1{|G|}\sum_{g\in G}U_g,
  \qquad
  \eta_{Y/G}(s)=\frac1{|G|}\sum_{g\in G}
  \Tr\!\left(U_gD_Y|D_Y|^{-s-1}\right).
  \label{eq:general-projected-eta}
 \end{equation}
 If the internal flat connection is specified \(\rho:G\to H\), \(U_g\) contains both
 the geometric action and the holonomy \(\rho(g)\) of the gauge bundle. Its character therefore remains in
 the projected phase even when the internal curvature vanishes.

 \item \textbf{Equivariant \(K\)-theory localization of the symbol.}
 The principal symbol of the bulk chiral Dirac operator is the bundle map
 \begin{equation}
  \sigma(D_{X,E}^+)(x,\xi)
  =\mathrm i\,c_x(\xi)\otimes\one_E:
  S_{X,x}^+\otimes E_x\longrightarrow S_{X,x}^-\otimes E_x.
  \label{eq:general-principal-symbol}
 \end{equation}
 It is invertible for \(\xi\ne0\), so it defines a compactly supported class
 \([\sigma(D_{X,E}^+)]\in K_{G,c}^0(T^*X)\). After localization at the
 character of \(g\), the compactly supported equivariant \(K\)-group over
 \(T^*(X\setminus X^g)\) vanishes. Hence the localized symbol class is
 supported near \(X^g\); this is the precise reason that its index is pushed
 forward from the fixed set rather than from the whole of \(X\).
 For a fixed component
 \(F\subset X^g\), the relevant embedding is the cotangent-space embedding
 \(\jmath:T^*F\hookrightarrow T^*X\), whose normal bundle is
 \(N_F\oplus N_F^*\simeq N_F\otimes\C\). The \(K\)-theory
 self-intersection formula reads
 \begin{equation}
  \jmath^*\jmath_!(a)
  =a\otimes\lambda_{-1}(N_{F,\C}^*),
  \qquad
  \lambda_{-1}(V^*)=\sum_q(-1)^q\Lambda^qV^*.
  \label{eq:general-K-self-intersection}
 \end{equation}
 For one complex normal line \(L\), this is simply the virtual class
 \(1-L^*\) obtained by restricting the Koszul complex
 \(L^*\xrightarrow{z}\one\) to its zero section. Since the normal bundle
 has no \(g\)-fixed vector, the Euler class is invertible after localization
 at \(g\), and hence
 \begin{equation}
  [\sigma(D_{X,E}^+)]_{(g)}
  =\sum_{F\subset X^g}\jmath_!\!\left(
  \frac{\jmath^*[\sigma(D_{X,E}^+)]}
  {\lambda_{-1}(N_{F,\C}^*)}\right).
  \label{eq:general-K-localization}
 \end{equation}
 This algebraic localization is the reason why a non-identity group element
 contributes only on its fixed set \cite{AtiyahSegalII,AtiyahBott}.

 \item \textbf{From the self-intersection factor to the Donnelly density.}
 The decomposition
 \(S_X|_F\simeq S_F\mathbin{\widehat\otimes}S(N_F)\) gives
 \begin{equation}
  \jmath^*[\sigma(D_{X,E}^+)]
  =[\sigma(D_{F,E|_F}^+)]\otimes
  \bigl([S^+(N_F)]-[S^-(N_F)]\bigr).
  \label{eq:general-restricted-symbol}
 \end{equation}
 Thus the fixed-point normal \(K\)-class is the normal-spinor numerator of
 the restricted Dirac symbol divided by the self-intersection Euler class
 of the cotangent bundle. Taking the equivariant Chern character gives
 \begin{equation}
  \boxed{
  \mathcal K_{\widetilde g}(N_F)
  =\frac{\STr_{S(N_F)}\!\left(\widetilde g\,e^{-\mathsf R_S}\right)}
  {\det_{N_F\otimes\C}\!\left(1-g\,e^{-\mathsf R_N}\right)}
  =\varepsilon(\widetilde g)\prod_{\alpha=1}^{q}
  \frac1{2\sinh\!\left((x_\alpha+\mathrm i\theta_\alpha)/2\right)} .}
  \label{eq:general-normal-factor}
 \end{equation}
 Here \(\rank_{\Rset}N_F=2q\), \(\widetilde g\) is the lift to the normal
 spinors, and \(\varepsilon(\widetilde g)\) keeps track of the orientation and
 spin-lift convention. The characteristic expression used below as the
 candidate local anomaly class on one fixed component is
 \begin{equation}
  I_{g,F}=\left[\widehat A(TF)\,
  \mathcal K_{\widetilde g}(N_F)\,\ch_g(E|_F)\right]_{d+2},
  \qquad
  \ch_g(E)=\Tr_{E|_{X^g}}\!\left(g\,e^{\mathrm iF_E/(2\pi)}\right).
  \label{eq:general-fixed-density}
 \end{equation}
 Since the fixed locus in every \(D^4\) fiber is the origin, the fixed
component is a section isomorphic to \(W_6\), and the internal
fixed-fiber pushforward reduces to the identity. As usual in
anomaly-polynomial computations, the resulting characteristic
expression is understood as a universal class. Equivalently, it may be
evaluated on an auxiliary \(W_8\), for which the fixed component of
\(W_8\times D^4\) is \(W_8\times\{0\}\) and the ordinary fixed-point
integral selects its degree-eight part. This accounts for the
degree-eight bracket in equation~\eqref{eq:corner-single-sector-review}.
Its direct identification with the projected spectral problem on the
physical nine-dimensional boundary is the prescription discussed
below. We next determine the Ho\v{r}ava--Witten virtual fermion bundle
and evaluate the finite sum in
equation~\eqref{eq:corner-group-average-review}.
 Equivariant \(K\)-theory fixes this bulk density. On a manifold with
 boundary, the APS spectral projector supplies the reduced equivariant
 \(\eta\)-invariant, and Donnelly's formula is
 \begin{equation}
  \ind_g(D_{X,E}^+)
  =\sum_{F\subset X^g}\int_F\mathcal I_{g,F}(D_{X,E}^+)
  -\xi_g(D_{Y,E}),
  \qquad \xi_g=\frac{\eta_g+h_g}{2}.
  \label{eq:general-donnelly}
 \end{equation}
 Appendix~\ref{app:k-theory-denominator} explains the origin of the numerator,
 denominator, and rotation-angle form of
 equation~\eqref{eq:general-normal-factor} in more detail. The polynomial \(I_{g,F}\) is
 only the contribution of one group element on one fixed component.

 \item \textbf{Average over the group.}
 For a common singular component \(F\), the projected non-identity
 fixed-point class is
 \begin{equation}
  I_{G,F}:=\frac1{|G|}
  \sum_{\substack{g\in G,\ g\ne1\\F\subset X^g}}I_{g,F}.
  \label{eq:general-group-summed-fixed-density}
 \end{equation}
 Including the identity sector gives the formal projected one-loop
 characteristic class
 \begin{equation}
  I_G^{\rm 1-loop}=\frac1{|G|}I_{1,X}+\sum_F I_{G,F}.
  \label{eq:general-full-orbifold-anomaly}
 \end{equation}
 The first term is the ordinary index density on the covering space, while the second
 contains the non-identity fixed-point corrections.
\end{itemize}

In general the fixed-point density must still be integrated along its
internal space. In the orbi-instanton theory we will consider below, the fixed component is
the external spacetime itself, \(F=W_6\), so \(F\to W_6\) is the identity and
there is no additional pushforward.

These four steps take the projected boundary
spectrum to the bulk fixed-point density. The next subsection 
specializes this equivariant \(K\)-theory prescription to the complex two-dimensional
normal bundle of the \(A_{k-1}\) geometry.
\subsubsection{\texorpdfstring{The normal contribution on \(\C^2/\Z_k\)}
{The normal contribution on C2/Zk}}
\label{sec:A-type-normal-contribution}

For the \(A_{k-1}\)-type ALE singularity, the general prescription is explicit. The
wall fermions propagate on
\[
 X=W_6\times D^4,\qquad Y=W_6\times S^3,
\]
and \(\Z_k\) acts on \(D^4\subset\C^2\). A non-identity element acts freely
on \(S^3\) but fixes the origin of \(D^4\), so the fixed set is \(W_6\) and
its real rank-four normal bundle is
\[
 N\simeq\Rset^4=\Rset^2_+\oplus\Rset^2_-\simeq\C^2.
\]
For \(g^j\in\Z_k\),
\begin{equation}
 g^j:(z_1,z_2)\longmapsto
 (e^{\mathrm i\theta_j}z_1,e^{-\mathrm i\theta_j}z_2),
 \qquad \theta_j=\frac{2\pi j}{k},\qquad j=1,\ldots,k-1.
 \label{eq:normal-angle-definition}
\end{equation}
The number \(\theta_j\) is the discrete \(SO(2)\) rotation angle on a real
normal two-plane, a zero-form defined modulo \(2\pi\). A spin lift sees the half-angle and is defined modulo \(4\pi\).
We use the standard lift compatible with
\eqref{eq:normal-angle-definition}. The paired weights
\((+\theta_j,-\theta_j)\) make the final four-dimensional factor invariant
under \(j\leftrightarrow k-j\). Appendix~\ref{app:normal-angle-conventions}
gives the conventions in detail.
Throughout this paper, \(L(k,1)\) denotes the quotient of \(S^3\) by
the action in equation~\eqref{eq:normal-angle-definition}, equipped
with the boundary orientation inherited from \(D^4\).

The flat equivariant Chern character of the cotangent-bundle
self-intersection Euler class in the denominator is
\begin{equation}
 \det_{N\otimes\mathbb C}(1-g^j)
 =16\sin^4(\theta_j/2),
 \label{eq:normal-euler-4d}
\end{equation}
whereas the normal-spinor virtual class in the restricted Dirac symbol in the numerator gives
\begin{equation}
 \STr_{S(\Rset^4)}(g^j)=4\sin^2(\theta_j/2).
 \label{eq:clifford-supertrace-4d}
\end{equation}
Combining them gives equation~\eqref{eq:general-normal-factor} in the form
\begin{equation}
 \frac{1}{4\sin^2(\theta_j/2)}.
 \label{eq:flat-spin-denominator}
\end{equation}
Thus the square root of the \(\sin^{-4}(\theta_j/2)\) from the self-intersection
Euler class is cancelled by the Clifford supertrace of the normal-spinor
virtual class. If \(r\) is the formal normal curvature root, with
\begin{equation}
 r^2=-c_2(R),
 \label{eq:general-R-root}
\end{equation}
the full normal factor is
obtained by evaluating the numerator and denominator of
equation~\eqref{eq:general-normal-factor} separately. Define
\begin{equation}
 Y_{j,+}=r+\mathrm i\theta_j,
 \qquad Y_{j,-}=r-\mathrm i\theta_j.
 \label{eq:normal-equivariant-roots}
\end{equation}
Then the curvature-refined normal-spinor numerator and self-intersection
Euler denominator are
\begin{align}
 \mathcal N_j(r)
 &:=\STr_{S(N)}\!\left(\widetilde g^j e^{-\mathsf R_S}\right)
 =4\sinh\!\left(\frac{Y_{j,+}}2\right)
   \sinh\!\left(\frac{Y_{j,-}}2\right),
 \label{eq:normal-spin-numerator}\\
 \mathcal D_j(r)
 &:=\det_{N\otimes\mathbb C}\!\left(1-g^j e^{-\mathsf R_N}\right)
 =16\sinh^2\!\left(\frac{Y_{j,+}}2\right)
    \sinh^2\!\left(\frac{Y_{j,-}}2\right).
 \label{eq:normal-vector-denominator}
\end{align}
At \(r=0\), these reduce respectively to
equations~\eqref{eq:clifford-supertrace-4d} and
\eqref{eq:normal-euler-4d}. Their ratio gives the full normal factor,
\begin{equation}
 \boxed{
 K_j(r)=\frac{\mathcal N_j(r)}{\mathcal D_j(r)}=
 \frac1{4\sinh((r+\mathrm i\theta_j)/2)
 \sinh((r-\mathrm i\theta_j)/2)}
 =\frac1{2(\cosh r-\cos\theta_j)} .}
 \label{eq:Kkernel}
\end{equation}
At \(r=0\), it reduces to equation~\eqref{eq:flat-spin-denominator}. A single
group element contributes
\begin{equation}
 I_{g^j,W}(E)=
 \left[\widehat A(TW)K_j(r)
 \ch_{g^j,\rho(g^j)}(E|_{W_6})\right]_8.
 \label{eq:K-fixed-density}
\end{equation}
This \(I_{g^j,W}\) is only the \(j\)-th twisted sector. The complete
non-identity wall anomaly is (notice that the summation starts from $j = 1$ instead of $j = 0$)
\begin{equation}
 \boxed{
 I_{\Z_k,W}(E;\rho)
 =\frac1k\sum_{j=1}^{k-1}I_{g^j,W}(E)
 =\left[\widehat A(TW)\frac1k\sum_{j=1}^{k-1}
 K_j(r)\ch_{g^j,\rho(g^j)}(E|_{W_6})\right]_8 .}
 \label{eq:general-ALE-fixed-inflow}
\end{equation}

To further expand the expression $K_j(r)$, whose denominator involves the formal sum of the zero-form $\theta_j$ and the two-form $r$, we define
\begin{equation}
 D_j=2-2\cos\theta_j=4\sin^2\frac{\theta_j}{2}.
 \label{eq:Dj-def}
\end{equation}
Using \(r^2=-c_2(R)\), we get $K_j(r)$ as a formal sum organized by powers of $c_2(R)$:
\begin{equation}
 K_j(r)=D_j^{-1}+c_2(R) D_j^{-2}
 +c_2(R)^2\left(D_j^{-3}-\frac1{12}D_j^{-2}\right)+O(c_2(R)^3).
 \label{eq:KexpandR}
\end{equation}
\section{Anomaly inflow of A-type orbi-instanton theories }
\label{sec:inflow}

We now apply the fixed-point framework of
Section~\ref{sec:orbifold-aps} to the six-dimensional SCFTs
constructed from \(N\) M5-branes probing an M9 wall and a transverse
A-type ALE singularity. Denote the non-identity ALE fixed-point class on the M9 wall
by \(c_{\rm wall}(\rho)\). Motivated by the decomposition proposed in
\cite{MOTZ}, define
\begin{equation}
 I_8^{\rm cand}(\rho)
 :=I_8^{\rm naive}(Q(\rho))+c_{\rm wall}(\rho).
 \label{eq:three-part-decomposition}
\end{equation}
The first term contains the established M5/Ho\v{r}ava--Witten and
ALE/Ho\v{r}ava--Witten contributions. We compute the second term for
an arbitrary \(\rho:\Z_k\to E_8\) from the conjectural family
fixed-point correspondence described above and establish the
polynomial identity
\[
 I_8^{\rm cand}(\rho)=I_8^{\rm TB}(\rho).
\]
This is a stringent test of the orbifold fixed-point framework, but
does not by itself construct the complete M-theory anomaly-line
trivialization.

\subsection{Tensor-branch anomalies and the conjectural \texorpdfstring{\(I_8(\rho)\)}{I8(rho)}}
\label{sec:tensor-review}

\subsubsection{Six-dimensional tensor-branch anomalies}

\paragraph{From chiral fermions to an eight-form.}
The local anomaly of a six-dimensional chiral theory is encoded by an
eight-form \(I_8\). For gauge transformations or diffeomorphisms, the descent
relations are
\begin{equation}
 I_8=\dd I_7^{(0)},\qquad
 \delta I_7^{(0)}=\dd I_6^{(1)},\qquad
 \delta\Gamma=2\pi\mathrm i\int_{W_6}I_6^{(1)}.
 \label{eq:descent-review}
\end{equation}
The one-loop anomaly of a complex Weyl fermion is the index density
\begin{equation}
 I_8^{\rm fermion}
 =\epsilon\left[\widehat A(TW)\ch_{R}(F)\right]_8,
 \label{eq:fermion-index-density}
\end{equation}
where \(\epsilon=\pm1\) is fixed by the chirality convention. A complex Weyl
fermion must be distinguished from a chiral fermion obeying a reality (Majorana)
condition. In the latter case the quadratic action is antisymmetric, the
Grassmann integral gives a Pfaffian so that
\(\operatorname{Pf}(D)^2=\det D\). Its logarithm and anomaly are therefore
one half of those of a complex Weyl fermion. Fermions in six-dimensional
\((1,0)\) multiplets obey a symplectic Majorana--Weyl condition. Our
`real-fermion convention'' counts these physical degrees of freedom
\cite{AGW,OSTY6d}. Expanding \(\widehat A\) and \(\ch_R\) gives gravitational
terms spanned by \(p_1^2,p_2\), gauge terms containing \(\Tr_R F^4\), and
mixed terms \(p_1\Tr_R F^2\). Here \(F\) may be the field strength of a
dynamical gauge symmetry, \(SU(2)_R\), or a flavor symmetry. Summing over all
fields and their representations gives the tensor-branch one-loop anomaly
\(I_8^{\rm 1-loop}\) \cite{OSTY6d}.

\paragraph{Green--Schwarz term.}
Let \(H_i\) be the self-dual field strengths on the tensor branch, with
string-charge pairing \(\Omega_{ij}\). If their modified Bianchi identities
are \(\dd H_i=X_i\), the total anomaly including the Green--Schwarz
contribution is
\begin{equation}
 I_8^{\rm TB}=I_8^{\rm 1-loop}
 +\frac12\,\Omega^{ij}X_iX_j,
 \qquad
 X_i=a_ic_2(R)+b_i p_1(TW)+\sum_A c_{iA}\Tr F_A^2 .
 \label{eq:TB-general}
\end{equation}
In practice, the irreducible \(\Tr F^4\) term of every dynamical gauge factor
must cancel in the one-loop result. This fixes the pairing between tensor
multiplets and gauge factors. The remaining
\((\Tr F^2)^2\), \(c_2(R)\Tr F^2\), and \(p_1\Tr F^2\) terms must factorize as
\(\Omega^{ij}X_iX_j/2\). These conditions determine the \(X_i\). Eliminating
the characteristic classes of the dynamical gauge fields leaves the
eight-form \(I_8^{\rm TB}\) built from \(SU(2)_R\), gravity, and flavor
backgrounds.

A tensor-multiplet expectation value breaks conformal symmetry but preserves
these internal symmetries. The ultraviolet SCFT and the tensor-branch
effective theory must therefore have the same anomaly. The algorithm of
\cite{MOTZ} uses the Kac labels to determine the nodes, gauge groups, and
matter representations, and then applies the cancellation and factorization
conditions node by node. The F-theory tensor branch is case-dependent,
but the final anomaly has a uniform Kac-polynomial form. This strongly
suggests a computation that does not depend on the detailed
tensor-branch presentation. Our non-identity ALE fixed-point
computation on the M9 wall reproduces the Kac-label-dependent remainder in such a uniform
form and can retain characteristic classes for the wall-centralizer
background.

\subsubsection{The conjectural \texorpdfstring{\(I_8(\rho)\)}{I8(rho)} formula}

Consider the six-dimensional \((1,0)\) SCFT obtained from \(N\) M5-branes
near the Ho\v{r}ava--Witten \(E_8\) wall and a transverse \(A_{k-1}\)
singularity. The flat \(E_8\) connection at infinity is specified by
\[
  \rho:\Z_k\longrightarrow E_8
\]
Conjugacy classes of finite-order elements in a compact Lie group are
described by Kac coordinates on the affine Dynkin diagram
\cite{KacFiniteOrder}. \cite{MOTZ} applied this description to an
arbitrary \(\rho\) and gave an algorithm for constructing the tensor branch
and its anomaly. Choose a maximal torus of \(E_8\) and write
\begin{equation}
  h_\rho=\rho(1)=\exp\!\left(\frac{2\pi i}{k}w\right),
  \qquad
  w=\sum_{i=1}^{8}n_i\omega_i^\vee .
  \label{eq:kac-holonomy}
\end{equation}
where the \(n_i\), together with the affine-node label, obey the level-\(k\)
Kac constraint. For each positive root \(\alpha>0\), distinguish the
unreduced pairing from its residue modulo \(k\):
\begin{equation}
  \widetilde a_\alpha:=\langle w,\alpha\rangle,
  \qquad
  a_\alpha:=[\widetilde a_\alpha]_k,
  \qquad 0\leq a_\alpha<k.
  \label{eq:aalpha}
\end{equation}
For the dominant Kac representative, \(0\leq\widetilde a_\alpha\leq k\).
It is therefore consistent to set
\begin{equation}
 s_\alpha
 :=\widetilde a_\alpha(k-\widetilde a_\alpha)
 =a_\alpha(k-a_\alpha).
 \label{eq:s-alpha-unreduced}
\end{equation}
The equality also holds at the endpoint \(\widetilde a_\alpha=k\), where
the residue is \(a_\alpha=0\). Group characters depend only on
\(a_\alpha\), whereas the Kac power sums below use the unreduced
\(\widetilde a_\alpha\).

\paragraph{Conjectural formula.}
Comparing tensor-branch anomalies with the naive M-theory inflow, \cite{MOTZ} found
\begin{equation}
 I_8^{\rm TB}=I_8^{\rm inflow,naive}(Q)+c(\rho),
 \qquad
 Q=N+\frac12\left(k+\frac1k-\frac{\langle w,w\rangle}{k}\right),
 \label{eq:MOTZ-Q}
\end{equation}
Here the \textit{naive inflow} term includes an E-string contribution (with
the ALE singularity omitted) and a conformal-matter contribution (with the M9
wall omitted). The remainder \(c(\rho)\), viewed as the codimension-five
corner term, can be organized by its scaling with \(k\):
\begin{equation}
 c(\rho)=\frac1k\bigl(P_0+P_2+P_4+P_6\bigr)
 +\frac12 I_{\rm vector}^{\rm free}.
 \label{eq:MOTZ-inductive}
\end{equation}
Here \(P_{2m}\) is homogeneous of degree \(2m\) in the Kac data, with \(w\)
assigned degree one in \(k\). Once again, we reproduce the explicit expressions from
\cite{MOTZ}. Let \(\Delta_+\) be the set of positive roots and define the
\(E_8\) Weyl vector by
\begin{equation}
 \rho_{\rm W}:=\frac12\sum_{\alpha\in\Delta_+}\alpha.
 \label{eq:MOTZ-Weyl-vector}
\end{equation}
We write \(p_i=p_i(TW_6)\), and
\(c_2(R):=\Tr F_R^2\) for the second Chern class of \(SU(2)_R\). Geometrically,
\(SU(2)_R\) is one factor in the structure group
\(SO(4)\cong SU(2)_L\times SU(2)_R\) of the normal bundle of \(W_6\) in the
ten-dimensional wall. The homogeneous components of \(c(\rho)\) are
\begin{align}
 P_0={}&\frac1{384}
 \bigl(-88c_2(R)^2+32c_2(R) p_1-5p_1^2+4p_2\bigr),
 \label{eq:MOTZ-P0}\\
 P_2={}&\frac{k^2}{11520}
 \bigl(2512c_2(R)^2-760c_2(R) p_1+157p_1^2-124p_2\bigr)\notag\\
 &+\frac{15\langle w,w\rangle-k\langle w,\rho_{\rm W}\rangle}{5760}
 \bigl(112c_2(R)^2-40c_2(R) p_1+7p_1^2-4p_2\bigr),
 \label{eq:MOTZ-P2}\\
 P_4={}&-\frac1{288}
 \left(
 9\langle w,w\rangle^2+15k^2\langle w,w\rangle-2k^4
 -k\sum_{\alpha\in\Delta_+}\langle w,\alpha\rangle^3
 \right)\notag\\[-1mm]
 &\hspace{29mm}\times\bigl(4c_2(R)^2-c_2(R) p_1\bigr),
 \label{eq:MOTZ-P4}\\
 P_6={}&\frac1{240}
 \left(
 5\langle w,w\rangle^3+15k^2\langle w,w\rangle^2
 -5k^4\langle w,w\rangle+k^6
 -k\sum_{\alpha\in\Delta_+}\langle w,\alpha\rangle^5
 \right)c_2(R)^2,
 \label{eq:MOTZ-P6}\\
 I_{\rm vector}^{\rm free}={}&\frac1{5760}
 \bigl(-240c_2(R)^2-120c_2(R) p_1-7p_1^2+4p_2\bigr).
 \label{eq:MOTZ-free-vector}
\end{align}
Equations~\eqref{eq:MOTZ-P0}--\eqref{eq:MOTZ-P6} are the expressions
that we will reproduce from the non-identity ALE fixed-point class on the M9 wall.

\paragraph{The problem.}
Equation~\eqref{eq:MOTZ-inductive} was conjectured from a large set of
tensor-branch computations using their F-theory descriptions. Its
uniform dependence on the Kac data through \(w\) calls for a
construction that does not proceed quiver by quiver. Reproducing
equation~\eqref{eq:MOTZ-inductive} for every A-type homomorphism is
the principal algebraic test of the orbifold fixed-point prescription
used in this paper. Extending this prescription to D- and E-type
orbifolds, or promoting it to a complete M-theory anomaly-line
construction, requires additional geometric input.

\subsection{Known inflow contributions and ALE fixed-point term on M9}

\label{sec:equivariant-inflow}
\label{sec:wall-fixed-sum}

\subsubsection{Reviewing Ho\v{r}ava--Witten and ALE contributions}

\paragraph{Local geometry}
Let \(W_6\) be the six-dimensional worldvolume. We use the
characteristic-class conventions stated below. Near the wall and the ALE
singularity, the local eleven-dimensional model is
\begin{equation}
 Y_{11}^{\rm loc}=W_6\times\Rset_+\times\C^2/\Z_k,
 \qquad
 X_{10}^{\rm wall}=W_6\times\C^2/\Z_k .
 \label{eq:local-geometry}
\end{equation}
Excising \(B^4/\Z_k\) around the zero section produces the boundary
\(W_6\times L(k,1)\). There is one Lens space \(L(k,1)\) over each
\(x\in W_6\), so the total space forms a fibration
\begin{equation}
 L(k,1)\longrightarrow W_6\times L(k,1)\longrightarrow W_6.
 \label{eq:lens-family}
\end{equation}
Its nontrivial topological data are partly recorded by the \(R\)-symmetry
anomaly. At the fixed zero section,
\begin{equation}
 T\!\left(W_6\times D^4\right)|_{W_6}\cong TW\oplus N,
 \qquad
 r^2=-c_2(R),\quad p_i=p_i(TW).
 \label{eq:R-convention}
\end{equation}
where the rank-four real bundle \(N\) is the normal bundle. The variable
\(r\) is the \(SU(2)_R\) Chern root supplied by the splitting
principle. It satisfies \(r^2=-c_2(R)\) and is used only to expand
characteristic classes.

\paragraph{Trace conventions.}
We write \(\Tr_{248}\) for the trace in the adjoint representation of
\(E_8\) and
\(\Tr_{E_8}:=\frac1{30}\Tr_{248}\) for the quadratic form normalized
as in the Ho\v{r}ava--Witten Bianchi identity. For \(SU(k)\),
\(\tr\) denotes the trace in the fundamental representation. The
notation \(c_2(R)=\Tr F_R^2\) follows the six-dimensional anomaly
convention of \cite{MOTZ}. All curvatures entering Chern characters
are normalized by the same \(2\pi\) convention.

\paragraph{Known inflow contribution}

We first recall the ``naive inflow" terms of \cite{MOTZ}. Let \(Q\)
be the effective M5-brane charge. In a background for the
wall-centralizer symmetry, the known contribution is
\begin{align}
 I_8^{\rm naive}(Q;F_H)={}&
 \frac{Q^3k^2}{6}c_2(R)^2
 -\frac{Q^2k}{2}c_2(R)I_4(F_H)\notag\\
 &+Q\left(\frac12I_4(F_H)^2-\mathcal I_8\right)
 +\bigl(I_4(F_H)-Qkc_2(R)\bigr)J_4\notag\\
 &-\frac12 I^{\rm vec}\!\left(SU(k)\right),
 \label{eq:naive-inflow-original}
\end{align}
where
\begin{align}
 I_4(F_H)&=\frac14\left(p_1(TW)-2c_2(R)+\mathcal X_H\right),\notag\\
 I_4^{(0)}&:=I_4(0)
 =\frac14\left(p_1(TW)-2c_2(R)\right),\notag\\
 \mathcal X_H&:=\Tr_{E_8}F_H^2
 =\frac1{30}\Tr_{248}\mathsf F_H^2,\notag\\
 \mathcal I_8&=\frac1{48}\left[
 p_2(N)+p_2(TW)-\frac14\bigl(p_1(N)-p_1(TW)\bigr)^2\right],\notag\\
 J_4&=\frac1{48}\left(k-\frac1k\right)(4c_2(R)+p_1(TW))
 +\frac14\tr F_{SU(k)}^2 .
 \label{eq:naive-building-blocks}
\end{align}
The curvature \(F_H\) is embedded through
\(H_\rho\hookrightarrow E_8\), and \(\mathcal X_H\) uses the
\(E_8\) quadratic form induced by this embedding. Relative to the
unflavored expression, the additional terms are
\begin{equation}
 \begin{aligned}
 \Delta_HI_8^{\rm naive}
 :={}&I_8^{\rm naive}(Q;F_H)-I_8^{\rm naive}(Q;0)\\
 ={}&-\frac{Q^2k}{8}c_2(R)\mathcal X_H
 +\frac{Q}{4}I_4^{(0)}\mathcal X_H
 +\frac{Q}{32}\mathcal X_H^2
 +\frac14\mathcal X_HJ_4 .
 \end{aligned}
 \label{eq:naive-flavor-shift}
\end{equation}
Here \(I_4(F_H)\) is the four-form in the modified
Ho\v{r}ava--Witten Bianchi identity, \(\mathcal I_8\) is the
eight-form appearing in the eleven-dimensional one-loop coupling,
and \(J_4\) is the source carried by the \(A_{k-1}\) fixed plane.
Thus the known contribution already contains \(Q\)-dependent
wall-centralizer flavor terms. It does not, however, contain the
non-identity ALE fixed-point class of the fermions on the M9 wall.

\paragraph{Derivation of the known inflow contribution}
Equation~\eqref{eq:naive-inflow-original} combines two established
pieces: the eleven-dimensional M5/Ho\v{r}ava--Witten inflow
\cite{OSTeString} and the seven-dimensional inflow on the ALE fixed
plane \cite{OSTY6d}. The remaining \(N\)-independent polynomial is
the term that will be compared with the non-identity ALE fixed-point
computation on the M9 wall below.

For the eleven-dimensional part, let \(Y_{12}\) extend the
eleven-dimensional spacetime \(X_{11}\), and let \(G\) be the M-theory
four-form field strength. The relevant topological action is
\begin{equation}
 S_{\rm top}=2\pi\int_{Y_{12}}
 \left(\frac16G^3-G\mathcal I_8\right),
 \qquad \partial Y_{12}=X_{11}.
 \label{eq:M-topological-action}
\end{equation}
An M5-brane is a magnetic source for \(G\). Excise a tubular neighborhood of
its worldvolume and let \(e_4\) be the global angular form on the unit
four-sphere bundle of the normal bundle.\footnote{For
\(\pi:S(N)\to W_6\), the form \(e_4\) is closed and normalized by
\(\pi_*e_4=\int_{S^4}e_4=2\). It is a differential-form representative of
the angular class on each normal sphere and packages the magnetic flux in a
way that is globally defined over \(W_6\) \cite{FHMM}.} With a radial cutoff
\(\rho_{\rm b}\), the smoothed Bianchi identity can be written as
\begin{equation}
 \dd G=Q\,\dd\rho_{\rm b}\frac{e_4}{2},
 \qquad
 G'=G-Q\rho_{\rm b}\frac{e_4}{2},
 \qquad \dd G'=0
 \label{eq:smoothed-M5-flux}
\end{equation}
on the excised space. Substituting \(G'\) into
equation~\eqref{eq:M-topological-action} and integrating over the normal
four-sphere produces characteristic classes on \(W_6\).\footnote{The
Bott--Cattaneo pushforward identities
\(\pi_*(e_4^3)=2p_2(N)\) and \(\pi_*(e_4^2)=0\) turn the normal integral of
the cubic Chern--Simons term into the normal-bundle class \(p_2(N)\); terms
with fewer factors of \(e_4\) give the mixed contributions involving the
smooth background flux \cite{BottCattaneo,FHMM}.} The result is
\begin{equation}
 I_8^{\rm M5/HW}(Q)
 =\frac{Q^3k^2}{6}c_2(R)^2
 -\frac{Q^2k}{2}c_2(R) I_4(F_H)
 +Q\left(\frac12I_4(F_H)^2-\mathcal I_8\right).
 \label{eq:naive-M5-HW-block}
\end{equation}

The powers of \(k\) follow directly from the covering space. Let
\(p:\widetilde U\to U=\widetilde U/\Z_k\) be the \(k\)-fold cover away from
the ALE fixed point. For a compactly supported top form \(\omega\) on \(U\),
\begin{equation}
 \int_U\omega=\frac1k\int_{\widetilde U}p^*\omega,
 \qquad
 Q_{\rm cover}
 =\int_{\widetilde S^4}p^*G
 =k\int_{S^4/\Z_k}G=kQ ,
 \label{eq:cover-integrals}
\end{equation}
where the flux integral uses the normalization that defines the M5-brane
charge. Therefore
\begin{align}
 I_8^{\rm M5/HW}(Q)
 &=\frac1k I_{8,\rm cover}^{\rm M5/HW}(kQ)\notag\\
 &=\frac1k\left[
 \frac{(kQ)^3}{6}c_2(R)^2-\frac{(kQ)^2}{2}c_2(R) I_4(F_H)
 +(kQ)\left(\frac12I_4(F_H)^2-\mathcal I_8\right)\right].
 \label{eq:cover-naive-identity}
\end{align}
The factors \(k^2,k,1\) arise respectively from the cubic, quadratic, and
linear terms of the covering-space flux.

The \(A_{k-1}\) fixed plane \(X_7\) carries an \(SU(k)\) gauge theory in
seven dimensions and ends on \(W_6\). Reducing the local eleven-dimensional
couplings along the ALE directions gives
\begin{equation}
 2\pi\int_{X_7}C_{\rm bulk}\wedge J_4,\qquad
 J_4=\frac1{48}\left(k-\frac1k\right)(4c_2(R)+p_1(TW))
      +\frac14\tr F_{SU(k)}^2 .
 \label{eq:ALE-local-CS-origin}
\end{equation}
Here \(k-1/k\) is the integrated ALE curvature defect, while the gauge term
comes from the seven-dimensional \(SU(k)\) sector. Its boundary condition
contributes one half of a six-dimensional vector-multiplet anomaly, so
\begin{equation}
 I_8^{\rm ALE/HW}(Q)
 =(I_4(F_H)-Qkc_2(R))J_4-\frac12I^{\rm vec}(SU(k)).
 \label{eq:naive-ALE-HW-block}
\end{equation}
Adding equations~\eqref{eq:naive-M5-HW-block} and
\eqref{eq:naive-ALE-HW-block} gives
equation~\eqref{eq:naive-inflow-original}. The  M9-wall-fermion contribution at the
orbifold corner is still missing.

Finally, the effective M5-brane charge $Q$ is not an integer in general, but is the sum of a gauge-instanton charge
and a curvature-induced charge \cite{MOTZ}:
\begin{equation}
 Q=\int_{\C^2/\Z_k}\Tr_{E_8}F_{E_8}^2
   +\int_{\C^2/\Z_k}\frac{p_1(T_{\rm ALE})}{4}.
 \label{eq:Q-charge-geometric-origin}
\end{equation}
Here \(T_{\rm ALE}\) is the tangent bundle of the ALE four-space, and
the gauge trace follows the normalization stated above and in
\cite{MOTZ}. Substituting the Kac
holonomy gives equation~\eqref{eq:MOTZ-Q}. For the trivial homomorphism, set
\begin{equation}
 Q_0=N+\frac12\left(k+\frac1k\right).
 \label{eq:Q0}
\end{equation}
Since equation~\eqref{eq:naive-inflow-original} is at most cubic in \(Q\),
\(I_8^{\rm naive}\) contains all terms growing as \(N^3,N^2,N\) with the
number of M5-branes. The remaining \(N\)-independent polynomial is
denoted by \(c(\rho)\) in \cite{MOTZ}. We now compute the corresponding
non-identity ALE fixed-point class on the M9 wall and show that it equals this
polynomial.

\subsubsection{The non-identity ALE fixed-point contribution on the M9 wall}

The preceding section and Appendix~\ref{app:aps-to-donnelly}
determine the equivariant ALE fixed-point density of the M9-wall fermion
symbol. Interpreting its degree-eight component as the
lower-dimensional anomaly class would in principle require additional family-index input, and we adopt the fixed-point expression below as a physical
prescription. We test the prescription
against the tensor-branch anomaly polynomial. For \(g^j\in\Z_k\), the
four-dimensional normal-spinor virtual class and self-intersection
Euler class combine into
\[
 K_j(r)=\frac1{2(\cosh r-\cos\theta_j)},\qquad
 \theta_j=\frac{2\pi j}{k},
\]
while the geometric action and the Kac holonomy \(\rho(g^j)\) enter
the same equivariant Chern character. A single group element gives
one formal fixed-point class,
\begin{equation}
 I_{g^j,W}
 =\left[\widehat A(TW)K_j(r)
 \ch_{g^j,\rho(g^j)}(V_{\rm HW})\right]_8.
 \label{eq:corner-single-sector-review}
\end{equation}
The orbifold projection then averages over all group elements, with all non-identity twisted-sector terms contributing to the orbifold correction:
\begin{equation}
 I_{\Z_k,W}=\frac1k\sum_{j=1}^{k-1}I_{g^j,W}.
 \label{eq:corner-group-average-review}
\end{equation}
Since the fixed locus in every \(D^4\) fiber is the origin, the
fixed component is a section isomorphic to \(W_6\), and the internal
fixed-fiber pushforward reduces to the identity. This family
pushforward should not be confused with the numerical integration of
an eight-form over a fixed six-manifold. We next determine the
Ho\v{r}ava--Witten virtual fermion bundle and evaluate the finite sum
in equation~\eqref{eq:corner-group-average-review}.

\paragraph{Fermions on M9 and the equivariant Chern character}\leavevmode\par

We now apply equation~\eqref{eq:K-fixed-density} to the fields on the M9 wall. The
Ho\v{r}ava--Witten wall supports a ten-dimensional \(E_8\)
Majorana--Weyl gaugino. The bulk supergravity fermions restricted to the
half-space give the gauge-fixed Rarita--Schwinger--dilatino virtual
combination. In terms of complex vector bundles, we use
\cite{HoravaWitten,AGW,WittenGlobal}
\begin{equation}
 V_{\rm HW}
 =\frac12\Ad(E_8)+\frac14\bigl(T_\C X_{10}^{\rm wall}-2\bigr).
 \label{eq:HW-virtual}
\end{equation}
The factor \(1/2\) in \(\frac12\Ad(E_8)\) is the Pfaffian normalization of
a real chiral gaugino. In the second term,
\(T_\C X_{10}^{\rm wall}-2\) is the virtual tangent bundle obtained from the
Rarita--Schwinger operator after gauge fixing and ghost subtraction. The
overall factor \(1/4\) contains two factors of \(1/2\). The first again comes
from the real-fermion Pfaffian. The second comes from the Ho\v{r}ava--Witten
\(S^1/\Z_2\) projection: the bulk-gravitino anomaly localizes equally on the
two reflection fixed planes, so a single M9 boundary receives one half of
the ten-dimensional Rarita--Schwinger--dilatino contribution
\cite{HoravaWitten,AGW,WittenGlobal}.

Next, we put in the holonomy data into the equivariant Chern character. Let \(h_\rho=\rho(g)\in E_8\), and let
\(\mathsf F_{E_8}\) be the normalized curvature of the physical adjoint
bundle. By equation~\eqref{eq:general-fixed-density},
\begin{equation}
 \ch_{h_\rho^j}\!\left(\Ad(E_8)\right)
 =\Tr_{248}\!\left(h_\rho^j e^{\mathsf F_{E_8}}\right).
 \label{eq:E8-equiv-ch-full}
\end{equation}
If the complexified adjoint bundle is decomposed into \(h_\rho\)
eigenbundles, \(\Ad(E_8)_\C=\bigoplus_\lambda E_\lambda\), with
\(h_\rho|_{E_\lambda}=e^{2\pi\mathrm i\lambda}\), the same expression is
\begin{equation}
 \ch_{h_\rho^j}\!\left(\Ad(E_8)\right)
 =\sum_\lambda e^{2\pi\mathrm i j\lambda}\ch(E_\lambda).
 \label{eq:E8-equiv-ch-eigenspaces}
\end{equation}
For the moment, set the continuous flavor curvature to zero,
\(\mathsf F_{E_8}=0\). The equivariant Chern character then has only a
degree-zero part. To keep this specialization visible, define
\begin{equation}
 \ch_{h_\rho^j,\mathrm f}\!\left(\Ad(E_8)\right)
 :=\left.\ch_{h_\rho^j}\!\left(\Ad(E_8)\right)
 \right|_{\mathsf F_{E_8}=0}
 =\Tr_{248}(h_\rho^j)=\chi_{248}(h_\rho^j),
 \qquad
 \ch_{1,\mathrm f}\!\left(\Ad(E_8)\right)=248.
 \label{eq:chi-as-equiv-ch}
\end{equation}
The subscript \(\mathrm f\) means ``flat'': it denotes the degree-zero form
left after the continuous flavor curvature has been set to zero. We retain
the notation \(\ch_{h_\rho^j,\mathrm f}(\Ad(E_8))\) below to emphasize that
the adjoint character is the flat specialization of an equivariant Chern
character. A continuous flavor background is restored by retaining the
positive-degree terms in equation~\eqref{eq:E8-equiv-ch-full}.

The equivariant Chern character of the complexified normal bundle is
\begin{equation}
 \ch_{g^j}(N_\C)=4\cosh r\cos\theta_j.
 \label{eq:normal-ch}
\end{equation}
Therefore, the equivariant Chern character of the full virtual twisted gauge
bundle on the wall is
\begin{equation}
 \ch_{g^j,h_\rho^j}(V_{\rm HW})
 =\frac12\ch_{h_\rho^j,\mathrm f}\!\left(\Ad(E_8)\right)
 +\frac14\left[
 \ch(T_\C W)+4\cosh r\cos\theta_j-2\right].
 \label{eq:HW-equiv-ch}
\end{equation}
Equation~\eqref{eq:K-fixed-density} and the group average now give the
non-identity ALE fixed-point sum of the M9-wall fermions:
\begin{equation}
 c(\rho)
 :=\left[
 \widehat A(TW)\frac1k\sum_{j=1}^{k-1}K_j(r)
 \ch_{g^j,h_\rho^j}(V_{\rm HW})
 \right]_8 .
 \label{eq:wall-combined-master}
\end{equation}
Orbifold projection imposes the \(\Z_k\)-invariance condition by inserting
\begin{equation}
 \Pi_\rho=\frac1k\sum_{j=0}^{k-1}g^j\otimes h_\rho^j,
 \label{eq:projector}
\end{equation}
which correlates the orbifold action with the flat holonomy data. If \(D_{\widetilde L}\) is the boundary Dirac operator on the covering space \(S^3\) twisted by \(\Ad(E_8)\), then the spectral trace on the Lens space \(L(k,1)\) is
\begin{align}
 \eta_\rho(s)
 &=\Tr\!\left(\Pi_\rho D_{\widetilde L}
 |D_{\widetilde L}|^{-s-1}\right)\notag\\
 &=\frac1k\sum_{j=0}^{k-1}\eta_{g^j,h_\rho^j}(s).
 \label{eq:projected-eta-trace}
\end{align}
\paragraph{Three-dimensional versus nine-dimensional eta data.}
Strictly speaking, the APS boundary operator of the ten-dimensional
wall problem acts on the full nine-dimensional boundary
\[
 Y_9=W_6\times L(k,1),
\]
whereas equation~\eqref{eq:projected-eta-trace} uses the vertical
three-dimensional Dirac operator on the Lens-space fiber. For a
product metric in which the Lens-space fiber is rescaled by
\(\varepsilon^2\), the full boundary operator has the schematic
adiabatic decomposition
\begin{equation}
 D_{Y_9,\varepsilon}
 =D_{W_6}\widehat\otimes\one
 +\Gamma_{W_6}\widehat\otimes
 \varepsilon^{-1}D_{L/W}+O(\varepsilon).
 \label{eq:adiabatic-nine-three}
\end{equation}
The scalar equivariant eta-invariant of \(D_{L/W}\) is only the
degree-zero component of the equivariant eta-form of the boundary
family
\[
 L(k,1)\longrightarrow Y_9\longrightarrow W_6.
\]
Its positive-degree components contain the dependence on the external
gauge, tangent, and normal-bundle curvatures.

Consequently, the passage from the projected three-dimensional
spectral problem to the degree-eight expression
\eqref{eq:wall-combined-master} does not follow from the ordinary
numerical Donnelly theorem alone. It requires the equivariant
families APS theorem---the family refinement of the Donnelly
fixed-point formula---and an adiabatic-limit argument
\cite{BismutCheeger,Goette,LiuMa}. We expect such
a family-index analysis to justify the wall prescription. We use the relation
\begin{equation}
 ``g, \rho-\text{equivariant family }\eta \text{ invariant }''
 \quad\longleftrightarrow\quad
 \frac1k\sum_{j=1}^{k-1}
 K_j(r)\ch_{g^j,h_\rho^j}(V_{\rm HW})
 \label{eq:conjectural-family-correspondence}
\end{equation}
as a physical prescription and test
its degree-eight component against the tensor-branch anomaly
polynomial. The identity sector is separated as in
equation~\eqref{eq:intro-identity-sector-cancellation}, while the
remaining sectors \(j=1,\ldots,k-1\) give
equation~\eqref{eq:wall-combined-master}.

To expand the formula for an arbitrary homomorphism, use
\begin{align}
 \ch(T_\C W)
 &=6+p_1+\frac{p_1^2-2p_2}{12}+\cdots,\notag\\
 \widehat A(TW)
 &=1-\frac{p_1}{24}
 +\frac{7p_1^2-4p_2}{5760}+\cdots .
 \label{eq:Ahat-ch-main}
\end{align}
The bracket in the gravitational-superpartner numerator, including its
external factor \(1/4\), is then
\begin{equation}
 \begin{aligned}
 \frac14\!\left[\ch(T_\C W)+4\cosh r\cos\theta_j-2\right]
 ={}&1+\cos\theta_j+\frac{p_1}{4}
 -\frac{\cos\theta_j}{2}c_2(R)\\
 &+\frac{p_1^2-2p_2}{48}
 +\frac{\cos\theta_j}{24}c_2(R)^2+\cdots .
 \end{aligned}
 \label{eq:grav-equiv-ch-expand}
\end{equation}
Combining the degree-zero parts of the gaugino and the supergravity fermions, we define for later convenience
\begin{equation}
 b_j(\rho):=\frac12\ch_{h_\rho^j,\mathrm f}\!\left(\Ad(E_8)\right)
 +1+\cos\theta_j .
 \label{eq:bj-rho}
\end{equation}

\paragraph{Orbifold summation and trigonometric identities}

To proceed with the orbifold summation, define the non-identity root-of-unity sums
\begin{equation}
 C_m(k):=\frac1k\sum_{j=1}^{k-1}D_j^{-m},
 \qquad m=0,1,2,3 .
 \label{eq:Cm-def-main}
\end{equation}
Here \(j\) is the integer labeling the group element, \(\theta_j=2\pi j/k\).
A single \(D_j=4\sin^2(\pi j/k)\) still contains trigonometric functions, which can be simplified after summation. With \(\zeta_k=e^{2\pi\mathrm i/k}\),
\begin{equation}
 D_j=(1-\zeta_k^j)(1-\zeta_k^{-j}).
 \label{eq:D-root-unity}
\end{equation}
For composite \(k\), the nontrivial roots need not form a single Galois
orbit, but their full set is Galois invariant, so the sum is rational. Vieta
relations for \(P_k(z)=(z^k-1)/(z-1)\), or logarithmic derivatives at
\(z=1\), reduce these symmetric root-of-unity sums to rational polynomials
in \(k\). \footnote{For a quick check, the needed classical identities are
\(\sum_{j=1}^{k-1}\csc^2(\pi j/k)=(k^2-1)/3\),
\(\sum\csc^4(\pi j/k)=(k^2-1)(k^2+11)/45\), and
\(\sum\csc^6(\pi j/k)
=(k^2-1)(2k^4+23k^2+191)/945\).}The closed forms are
\begin{align}
 C_0&=\frac{k-1}{k},&
 C_1&=\frac{k^2-1}{12k},\notag\\
 C_2&=\frac{k^4+10k^2-11}{720k},&
 C_3&=\frac{2k^6+21k^4+168k^2-191}{60480k}.
 \label{eq:C123-main}
\end{align}
Since \(\cos\theta_j=1-D_j/2\), sums with one cosine reduce to
\begin{equation}
 \frac1k\sum_{j=1}^{k-1}\frac{\cos\theta_j}{D_j^m}
 =C_m-\frac12C_{m-1}.
 \label{eq:cos-Cm}
\end{equation}

We then define the weighted fixed-point sums
\begin{equation}
 B_m(\rho):=\frac1k\sum_{j=1}^{k-1}\frac{b_j(\rho)}{D_j^m},
 \qquad
 E_m(k):=\frac1k\sum_{j=1}^{k-1}\frac{\cos\theta_j}{D_j^m}
 =C_m-\frac12C_{m-1}.
 \label{eq:Bm-Em-def}
\end{equation}
Substituting equations~\eqref{eq:KexpandR}, \eqref{eq:Ahat-ch-main}, and
\eqref{eq:grav-equiv-ch-expand} directly into
equation~\eqref{eq:wall-combined-master}, the eight-form becomes
\begin{align}
 c(\rho)={}&
 c_2(R)^2\left(B_3-\frac1{12}B_2-\frac12E_2+\frac1{24}E_1\right)\notag\\
 &+c_2(R) p_1\left(\frac14C_2-\frac1{24}B_2+\frac1{48}E_1\right)\notag\\
 &+p_1^2\left(\frac1{96}C_1+\frac7{5760}B_1\right)
 -p_2\left(\frac1{24}C_1+\frac1{1440}B_1\right).
 \label{eq:c-rho-BCE}
\end{align}
This expression combines the gaugino and the gravitino contributions.

Let \(\Delta_+\) be the set of positive \(E_8\) roots. With
\(a_\alpha=[\widetilde a_\alpha]_k\), the root decomposition gives
\begin{equation}
 \frac12\ch_{h_\rho^j,\mathrm f}\!\left(\Ad(E_8)\right)
 =4+\sum_{\alpha>0}\cos(a_\alpha\theta_j),
 \qquad
 b_j(\rho)=5+\cos\theta_j+\sum_{\alpha>0}\cos(a_\alpha\theta_j).
 \label{eq:character-root}
\end{equation}
The constant \(4\) comes from the eight zero weights of the adjoint
representation, and each pair of roots \(\pm\alpha\) combines to give one cosine. The \(R\)-symmetry
still enters only through the normal root \(r\) in \(K_j(r)\), while \(p_1,p_2\) enter through \(\widehat A(TW)\).

\paragraph{Discrete Fourier sums.}

For a single root residue \(a\), define
\begin{equation}
 F_m(a;k):=\frac1k\sum_{j=1}^{k-1}
 \frac{\cos(a\theta_j)}{D_j^m},
 \qquad s=a(k-a).
 \label{eq:single-root-Fm}
\end{equation}
Performing a residue sum on the unit circle, we get
\begin{align}
 F_1(a;k)&=C_1-\frac{s}{2k},\notag\\
 F_2(a;k)&=C_2-\frac{s(s+2)}{24k},\notag\\
 F_3(a;k)&=C_3-\frac{s\,[2s^2+(k^2+15)s+24]}{1440k}.
 \label{eq:discrete-Fourier-main}
\end{align}
Complex conjugation exchanges \(a\) and \(k-a\), so the answer can depend
only on the symmetric combination \(s=a(k-a)\). Equation~\eqref{eq:character-root} then gives
\begin{equation}
 B_m(\rho)=5C_m+E_m+
 \sum_{\alpha>0}F_m(a_\alpha;k),
 \qquad m=1,2,3.
 \label{eq:Bm-from-Fm}
\end{equation}

\subsection{Summation over roots of \texorpdfstring{$E_8$}{E8} and comparison with \texorpdfstring{\cite{MOTZ}}{MOTZ}}
\label{sec:kac-comparison}

To compare equation~\eqref{eq:c-rho-BCE} with the Kac data in
section~\ref{sec:introduction}, we write the positive-root power sums in
our normalization. For even degree, define
\begin{equation}
 P_{2m}(w)
 :=\sum_{\alpha>0}\langle w,\alpha\rangle^{2m}
 =\frac12\Tr_{\mathbf{248}}\!\left((\operatorname{ad}w)^{2m}\right).
 \label{eq:E8-adjoint-moments}
\end{equation}
This is a homogeneous Weyl-invariant polynomial of degree \(2m\). A basis of the
Weyl-invariant of \(E_8\) (the Casimir invariants) are of degree
\[
 2,\ 8,\ 12,\ 14,\ 18,\ 20,\ 24,\ 30.
\]
Consequently, in degrees two, four, and six the only possibilities are
proportional to \(\langle w,w\rangle\), \(\langle w,w\rangle^2\), and
\(\langle w,w\rangle^3\), respectively. The constants are fixed by taking
\(w=\beta\) to be a root with \(\langle\beta,\beta\rangle=2\). For any given
\(\beta\), the 240 roots have inner products
\begin{equation}
 \begin{array}{c|ccccc}
 \langle\beta,\alpha\rangle&2&1&0&-1&-2\\ \hline
 \#\{\alpha\in\Delta\}&1&56&126&56&1
 \end{array}.
 \label{eq:E8-root-distribution}
\end{equation}
Their second, fourth, and sixth moments determine the coefficients below;
the linear identity follows directly from
\(\rho_{\rm W}=\frac12\sum_{\alpha>0}\alpha\):
\begin{align}
 \sum_{\alpha>0}\widetilde a_\alpha
 &=2\langle w,\rho_{\rm W}\rangle,&
 \sum_{\alpha>0}\widetilde a_\alpha^2
 &=30\langle w,w\rangle,\notag\\
 \sum_{\alpha>0}\widetilde a_\alpha^4
 &=18\langle w,w\rangle^2,&
 \sum_{\alpha>0}\widetilde a_\alpha^6
 &=15\langle w,w\rangle^3 .
 \label{eq:E8-powers-main}
\end{align}
Substitute
\(s_\alpha=\widetilde a_\alpha(k-\widetilde a_\alpha)\) into
equation~\eqref{eq:discrete-Fourier-main}, and use
equation~\eqref{eq:E8-powers-main} to perform the positive-root sum. Introduce the
three characteristic-class combinations
\begin{align}
 \mathcal A&=2512c_2(R)^2-760c_2(R) p_1+157p_1^2-124p_2,\notag\\
 \mathcal B&=112c_2(R)^2-40c_2(R) p_1+7p_1^2-4p_2,\notag\\
 \mathcal C&=4c_2(R)^2-c_2(R) p_1 .
 \label{eq:ABC}
\end{align}
After simplification, this agrees with equations (3.16)--(3.21) of \cite{MOTZ}:
\begin{align}
 c(\rho)={}&
 \frac{-88c_2(R)^2+32c_2(R) p_1-5p_1^2+4p_2}{384k}
 +\frac{k}{11520}\mathcal A\notag\\
 &+\frac{15\langle w,w\rangle-k\langle w,\rho_{\rm W}\rangle}{5760k}
 \mathcal B\notag\\
 &-\frac{1}{288k}\left(
 9\langle w,w\rangle^2+15k^2\langle w,w\rangle-2k^4
 -k\sum_{\alpha>0}\widetilde a_\alpha^3\right)\mathcal C\notag\\
 &+\frac{1}{240k}\left(
 5\langle w,w\rangle^3+15k^2\langle w,w\rangle^2
 -5k^4\langle w,w\rangle+k^6
 -k\sum_{\alpha>0}\widetilde a_\alpha^5\right)c_2(R)^2\notag\\
 &+\frac12I_{\rm vector}^{\rm free}.
 \label{eq:MOTZ-fully-expanded-again}
\end{align}
The three layers \(D_j^{-1},D_j^{-2},D_j^{-3}\) in the fixed-point
denominator therefore produce periodic polynomials of degree at most
\(s_\alpha,s_\alpha^2,s_\alpha^3\). Expanding
\(s_\alpha=\widetilde a_\alpha(k-\widetilde a_\alpha)\), the discrete
Fourier transform generates
the \(k,k^3,k^5\) layers and recovers the Kac-label-dependent polynomial of
section~\ref{sec:introduction}.

\subsection{Wall-centralizer flavor backgrounds}
\label{sec:flavor-main}

We first determine the dependence of the non-identity wall
fixed-point remainder on background fields for the symmetry selected
by the boundary holonomy. This remainder is not by itself the full
flavor anomaly: the known M5/Ho\v{r}ava--Witten contribution also
depends on the same background through \(I_4(F_H)\). For a
homomorphism \(\rho:\Z_k\to E_8\), let
\begin{equation}
 H_\rho=Z_{E_8}(h_\rho)
 \label{eq:flavor-centralizer}
\end{equation}
be the centralizer of \(h_\rho\), and denote its background curvature by
\(F_H\). The complexified adjoint representation decomposes into
\(\Z_k\)-weight spaces,
\begin{equation}
 \Ad(E_8)_\C=\bigoplus_{q=0}^{k-1}R_q\otimes\chi_q,
 \qquad h_\rho|_{\chi_q}=e^{2\pi\mathrm iq/k},
 \label{eq:flavor-decomposition}
\end{equation}
where each \(R_q\) is an \(H_\rho\)-representation. The equivariant Chern
character of the wall fermion in the \(j\)-th sector is therefore
\begin{equation}
 \ch_{h_\rho^j}\!\left(\Ad(E_8);F_H\right)
 =\sum_{q=0}^{k-1}e^{2\pi\mathrm i jq/k}\ch_{R_q}(F_H).
 \label{eq:flavor-equiv-ch}
\end{equation}
Substituting into the unified fixed-point formula gives the
non-identity wall remainder in a centralizer background,
\begin{equation}
 \begin{aligned}
 c_{\rm wall}(\rho;F_H)=\Biggl[&\widehat A(TW)\frac1k\sum_{j=1}^{k-1}K_j(r)\\[-2pt]
 &\times\left\{\frac12\ch_{h_\rho^j}(\Ad(E_8);F_H)
 +\frac14\left[\ch(T_\C W)+4\cosh r\cos\theta_j-2\right]\right\}
 \Biggr]_8 .
 \end{aligned}
 \label{eq:flavor-donnelly-master}
\end{equation}
Here \(F_H\) is regarded as an adjoint \(E_8\) curvature through
\(H_\rho\hookrightarrow E_8\); setting \(F_H=0\) recovers
equation~\eqref{eq:wall-combined-master}. To display the form-degree
bookkeeping, define
\begin{equation}
 \mathcal T_{2q}^{(m)}(\rho;F_H)
 :=\left[\frac1k\sum_{j=1}^{k-1}\frac{
 \ch_{g^j,h_\rho^j}(V_{\rm HW};F_H)}{D_j^m}\right]_{2q}.
 \label{eq:Tqm-def}
\end{equation}
Then
\begin{align}
 c_{\rm wall}(\rho;F_H)={}&\mathcal T_8^{(1)}
 +c_2(R)\mathcal T_4^{(2)}-\frac{p_1}{24}\mathcal T_4^{(1)}\notag\\
 &+c_2(R)^2\left(\mathcal T_0^{(3)}-\frac1{12}\mathcal T_0^{(2)}\right)
 -\frac{c_2(R) p_1}{24}\mathcal T_0^{(2)}
 +\frac{7p_1^2-4p_2}{5760}\mathcal T_0^{(1)}.
 \label{eq:flavor-degree-eight}
\end{align}
The complete perturbative anomaly polynomial in the
wall-centralizer background is therefore
\begin{equation}
 \boxed{
 I_8^{\rm TB}(\rho;F_H,F_R,F_{SU(k)})
 =I_8^{\rm naive}\!\left(
 Q(\rho);F_H,F_R,F_{SU(k)}\right)
 +c_{\rm wall}(\rho;F_H).}
 \label{eq:full-flavored-anomaly}
\end{equation}
When the ALE-end \(SU(k)\) background is not considered, one sets
\(F_{SU(k)}=0\). The discrete Fourier sums determine the
flavor-dependent coefficients of the wall remainder, whereas the
additional \(Q\)-dependent flavor terms are contained in
\(I_8^{\rm naive}\) through \(I_4(F_H)\).

\subsubsection{Centralizer decompositions and tensor branches}

The tensor-branch diagrams below display only the endpoint structure
needed to identify the indicated flavor action. The \(N\)-dependent
length of any repeated gauge-node plateau is suppressed; the diagrams
should not be read as complete quivers for arbitrary \(N\).

\paragraph{Trivial homomorphism.}
Here \(H_\rho=E_8\), and equation~\eqref{eq:flavor-equiv-ch} has no
nontrivial phase. In the absence of additional accidental enhancement, a
corresponding six-dimensional tensor-branch presentation is
\begin{equation}
 [E_8]\;1\;
 \overset{\mathfrak{su}(1)}{2}\;
 \overset{\mathfrak{su}(2)}{2}\;\cdots\;
 \overset{\mathfrak{su}(k-1)}{2}\;
 \underset{[U(1)]}{\overset{\mathfrak{su}(k)}{2}}\;
 \overset{\mathfrak{su}(k)}{2}\;[SU(k)] .
 \label{eq:trivial-tensor-branch-chain}
\end{equation}
The crucial point is that the last two consecutive \(\mathfrak{su}(k)\)
entries are both gauge nodes. The right flavor group \([SU(k)]\) acts only
on the rightmost gauge node, whereas \([U(1)]\) labels an independent
abelian flavor symmetry that is ``delocalized" in the sense of \cite{ApruzziU1}. The wall-centralizer background is
still only the left \([E_8]\) factor
\cite{HeckmanRudeliusReview,MOTZ}. With
\(\mathcal X_E=\Tr F_{E_8}^2\), the flavored wall
remainder is
\begin{align}
 c^{E_8}(1;F_{E_8})={}&\frac{k^2-1}{576k}\Tr_{248}\mathsf F^4\notag\\
 &+\left[\frac{k^4+10k^2-11}{2880k}c_2(R)
 -\frac{k^2-1}{1152k}p_1\right]\Tr_{248}\mathsf F^2,
 \label{eq:flavor-trivial-explicit}\\
 \Tr_{248}\mathsf F^2&=30\mathcal X_E,
 \qquad \Tr_{248}\mathsf F^4=9\mathcal X_E^2.
 \label{eq:E8-trace-conversion}
\end{align}
The three coefficients are respectively the quartic flavor term and its
mixed terms with \(c_2(R)\) and \(p_1\) in
equation~\eqref{eq:flavor-degree-eight}.

\paragraph{Nontrivial \(k=2\) homomorphism.}
For Kac label \((010000000)\) \cite{MOTZ}, the centralizer group is
\((E_7\times SU(2))/\Z_2\). Below we use representations of the
covering group \(E_7\times SU(2)\). A tensor-branch presentation that displays the two
ends is
\begin{equation}
 [E_7]\;1\;
 \underset{[N_f=2]}{\overset{\mathfrak{su}(2)}{2}}\;
 \overset{\mathfrak{su}(2)}{2}\;[SU(2)] .
 \label{eq:E7-tensor-branch-chain}
\end{equation}
The nonabelian flavor action carried by the left \([N_f=2]\) matter gives
the second \(SU(2)\) in the centralizer, whereas the right \([SU(2)]\)
comes from the ALE end and is kept distinct. Under \(E_7\times SU(2)\),
\begin{equation}
 \mathbf{248}=(\mathbf{133},\mathbf1)\oplus(\mathbf1,\mathbf3)
 \oplus(\mathbf{56},\mathbf2).
 \label{eq:E8-E7SU2-branching}
\end{equation}
The nontrivial group element acts with a minus sign only on the last
summand, so that
\begin{equation}
 \ch_{h_\rho}(\mathbf{248};F_{E_7},F_2)
 =\ch_{\mathbf{133}}(F_{E_7})+\ch_{\mathbf3}(F_2)
 -\ch_{\mathbf{56}}(F_{E_7})\ch_{\mathbf2}(F_2).
 \label{eq:E7SU2-equivariant-character}
\end{equation}
Substitution into equation~\eqref{eq:flavor-donnelly-master} gives all
\(E_7\times SU(2)\) background terms. With the flavor curvatures turned
off, it reduces to
\begin{equation}
 c_{E_7\times SU(2)}
 =\frac7{96}c_2(R)^2-\frac1{96}c_2(R) p_1
 +\frac1{320}p_1^2-\frac1{160}p_2.
 \label{eq:E7-c-explicit}
\end{equation}

\paragraph{The \(SO(14)\times U(1)_H\) centralizer at \(k=4\).}
For the Kac data \(1+1+2'\) discussed in
refs.~\cite{MOTZ,ApruzziU1}, the wall centralizer algebra is
\(\mathfrak{so}(14)\oplus\mathfrak u(1)_H\). The corresponding left
tensor-branch segment can be written as
\begin{equation}
 [SO(14)]\;
 \overset{\mathfrak{sp}(1)}{1}\;
 \overset{\mathfrak{su}(3)}{2}\;[SU(4)] .
 \label{eq:SO14-tensor-branch-chain}
\end{equation}
Here \(U(1)_H\) is the ABJ-anomaly-free linear combination acting on the
charged matter of this chain and belongs to the wall \(E_8\) centralizer;
the right \([SU(4)]\) is an ALE-end flavor symmetry and is not included in
\(F_H\). In the standard charge
normalization,
\begin{equation}
 \mathbf{248}=\mathbf{91}_0\oplus\mathbf1_0
 \oplus\mathbf{64}_{+1}\oplus\overline{\mathbf{64}}_{-1}
 \oplus\mathbf{14}_{+2}\oplus\mathbf{14}_{-2}.
 \label{eq:E8-SO14U1-branching}
\end{equation}
Writing \(\zeta_4=e^{2\pi\mathrm i/4}\), one obtains
\begin{align}
 \ch_{h_\rho^j}(\mathbf{248};F_{SO(14)},F_H)
 ={}&\ch_{\mathbf{91}_0}+1
 +\zeta_4^j\ch_{\mathbf{64}_{+1}}
 +\zeta_4^{-j}\ch_{\overline{\mathbf{64}}_{-1}}\notag\\
 &+(-1)^j\left(\ch_{\mathbf{14}_{+2}}+\ch_{\mathbf{14}_{-2}}\right).
 \label{eq:SO14U1-equivariant-character}
\end{align}
Each Chern character depends on the corresponding
\(SO(14)\times U(1)_H\) background. This decomposition can be inserted
directly into equation~\eqref{eq:flavor-donnelly-master}, giving the
prediction of our method for the wall-centralizer flavor anomaly. 

\section{Discussion}
\label{sec:discussion}

We have tested the orbifold fixed-point framework on A-type
orbi-instanton theories. The non-identity ALE fixed-point density on the M9 wall
reproduces the complete Kac-label-dependent remainder of the
unflavored eight-form. Retaining the positive-degree equivariant
Chern character determines the wall-centralizer flavor dependence of
this remainder, while the corresponding \(Q\)-dependent flavor terms
come from \(I_4(F_H)\) in the known contribution.

The computation tests the physical fixed-point
prescription used to obtain the perturbative local polynomial. We
expect this prescription to be correct because it reproduces the full
Kac-label-dependent result, and we expect a complete derivation for
the wall problem to require the equivariant family index theorem. In addition, we hope to extend the general anoamly inflow analysis so as to start from the full 11-dimensional bulk, e.g. comparing our analysis with that of \cite{Bah:2019jts}.

\paragraph{Limitationa regarding non-perturbative anomalies}
The analysis in this paper determines the perturbative eight-form anomaly polynomial. It does not determine torsion phases or the holonomy of
the anomaly line around a large gauge transformation or
diffeomorphism. A treatment of those questions would require the
complete M-theory \(C\)-field, Rarita--Schwinger, wall-gaugino, and
corner-boundary-condition system together with a gluing theorem. We
do not attempt that analysis here.

For clarity about conventions, the Rarita--Schwinger parity-anomaly
theory on a closed twelve-dimensional \(\mathrm{Pin}^+\) manifold is
written as
\begin{equation}
 \widehat\alpha_{\rm RS}^{\mathrm{Pin}^+}(W_{12})
 =
 \exp\left[
 2\pi\mathrm i\,
 \frac{\eta_{W_{12}}(TW_{12}-2)}4
 \right].
 \label{eq:discussion-RS-pinplus-anomaly}
\end{equation}
On the oriented spin subcategory it restricts to \cite{FreedMooreQuantum,FreedHopkinsMTheory} 
\begin{equation}
 \widehat\alpha_{\rm RS}^{\mathrm{Spin}}(W_{12})
 =
 (-1)^{\mathrm{RS}(W_{12})},
 \qquad
 \mathrm{RS}(W_{12})
 =
 \frac12\ind D^+_{W_{12},\,TW_{12,\C}-2}.
 \label{eq:discussion-RS-spin-anomaly}
\end{equation}

The local ten-dimensional wall anomaly in the real-fermion
convention is
\begin{equation}
 I_{12}^{\rm wall}
 =
 \left[
 \widehat A(TX_{10})
 \left\{
 \frac12\ch_{\mathbf{248}}(F_{E_8})
 +\frac14\left(\ch(T_{\C}X_{10})-2\right)
 \right\}
 \right]_{12}.
 \label{eq:discussion-HW-wall-anomaly}
\end{equation}
This is the local index class whose non-identity orbifold sectors are
used in the main computation. The cancellation of its identity
sector by the standard Ho\v{r}ava--Witten bulk system is assumed as in
equation~\eqref{eq:intro-identity-sector-cancellation}.

\paragraph{Geometric \(U(1)_L\) and \(SU(2)_L\).}
Section~\ref{sec:flavor-main} retains only flavor backgrounds from the M9 wall, which is directly
controlled by \(H_\rho=Z_{E_8}(h_\rho)\); it does not identify the ALE
normal-bundle isometry in advance with a current named on the tensor branch.
A useful direction for further generalization is to incorporate them into the normal bundle analysis, and then compare them with \cite{ApruzziU1} to obtain a general identification of these symmetries for the five infinite families of tensor-branch configurations in \cite{MOTZ}. This dictionary must simultaneously match the ALE-isometry charges of tensor-branch matter, the Green--Schwarz
couplings, and the normal-bundle background on the inflow side. \footnote{Many subscripts
\(L,R\) in ref.~\cite{ApruzziU1} label the left and right ends of a
tensor-branch diagram; they do not denote the factors
\(SU(2)_L\times SU(2)_R\) in the normal-rotation group \(SO(4)\).}

\paragraph{Atomic Higgsings of 6D SCFTs: induced flows.}
Another future direction is to study how anomaly polynomials change along the
induced flows generated by atomic Higgsings, to perform anomaly-matching
computations between the ultraviolet and infrared theories, and to investigate
the physical and geometric meaning of these matches
\cite{AtomicHiggsingsI,AtomicHiggsingsII}. The relation between
orbi-instantons and D-type class \(\mathcal S\) theories found in
\cite{OrbiInstantonsDClassS} suggests that such results may also help clarify
the role of D-type Kac labels.

\section*{Acknowledgements}
HYZ thanks Jiakang Bao, Simeon Hellerman, Noppadol Mekareeya, Ruben Minasian, and Yuji Tachikawa for helpful discussions. HYZ thanks Yuji Tachikawa for valuable comments which significantly improved our presentation. HYZ is supported by WPI Initiative, MEXT, Japan at Kavli IPMU, the University of Tokyo. HYZ gratefully acknowledges the hospitality and support of the Simons Center for Geometry and Physics, Stony Brook University, during the 2026 Simons Physics Summer Workshop, where part of this work was carried out.

\appendix

\section{Equivariant \texorpdfstring{\(K\)}{K}-theory fixed-point formula and the Donnelly boundary term}
\label{app:aps-to-donnelly}

In this appendix, we begin with the ordinary relation between an elliptic
operator, its principal symbol, and the \(K\)-theory pushforward that gives its
index. We then review the derivation of the equivariant spin--Dirac fixed-point density used
in the main text entirely within equivariant \(K\)-theory. Computing the
self-intersection formula on the cotangent bundle produces the denominator,
while the restriction of the Dirac symbol to the fixed set produces the
normal-spinor numerator. The boundary equivariant \(\eta\)-invariant depends
on the spectrum of the boundary Dirac operator. It cannot be derived from
ordinary topological \(K\)-theory alone and is supplied by Donnelly's
equivariant APS theorem. Equivariant localization and the self-intersection
formula are developed in \cite{AtiyahSegalII,AtiyahBott,AtiyahSingerI};
the boundary theorem is treated in
\cite{APS1,APS2,DonnellyPatodi,Donnelly,BravermanMaschler}.

\subsection{Objectives}

Let
\begin{equation}
 P:\Gamma(E_0)\longrightarrow\Gamma(E_1)
 \label{eq:app-general-elliptic-operator}
\end{equation}
be an elliptic differential operator of order \(m\) on a closed compact
manifold \(X\). In local coordinates,
\begin{equation}
 P=\sum_{|\alpha|\leq m}a_\alpha(x)\partial_x^\alpha,
 \qquad
 \sigma_m(P)(x,\xi)
 =\sum_{|\alpha|=m}a_\alpha(x)(\mathrm i\xi)^\alpha.
 \label{eq:app-general-principal-symbol}
\end{equation}
The second expression is the principal symbol. Ellipticity means that
\(\sigma_m(P)(x,\xi):(E_0)_x\to(E_1)_x\) is invertible for every
\(\xi\ne0\). If \(\pi:T^*X\to X\) is the projection, the two pulled-back
bundles and this isomorphism away from the zero section define
\begin{equation}
 [\sigma(P)]
 =\left[\pi^*E_0,\pi^*E_1,\sigma_m(P)\right]
 \in K_c^0(T^*X).
 \label{eq:app-nonequivariant-symbol-class}
\end{equation}
Lower-order deformations do not change this class or the Fredholm index. The
ordinary analytic index is
\begin{equation}
 \operatorname{ind}_{\rm an}(P)
 =\dim\ker P-\dim\operatorname{coker}P,
 \label{eq:app-ordinary-analytic-index}
\end{equation}
and the Atiyah--Singer theorem identifies it with the topological pushforward
\begin{equation}
 \boxed{
 \operatorname{ind}_{\rm an}(P)
 =\operatorname{ind}_{\rm top}\!\left([\sigma(P)]\right)
 =p_![\sigma(P)]
 \in K^0(\mathrm{pt})\cong\mathbb Z.}
 \label{eq:app-AS-symbol-pushforward}
\end{equation}
Here \(p:T^*X\to\mathrm{pt}\), and \(p_!\) denotes the \(K\)-theory Gysin
map constructed through a Thom map and Bott periodicity; it is not an ordinary
integral over the noncompact cotangent bundle. Applying the Chern character
and the Riemann--Roch formula to the twisted spin--Dirac symbol gives
\begin{equation}
 \operatorname{ind}(D_{X,E}^+)
 =\int_X\widehat A(TX)\operatorname{ch}(E).
 \label{eq:app-ordinary-dirac-index-density}
\end{equation}
These statements summarize the non-equivariant background. The equivariant
construction used below is the same pushforward with the ordinary \(K\)-group
replaced by equivariant \(K\)-theory and the target \(\mathbb Z\) replaced by
the representation ring \(R(G)\) \cite{AtiyahSingerI,AtiyahSegalII}.

On an even-dimensional spin manifold \(X\), the twisted chiral Dirac
operator is
\begin{equation}
 D_{X,E}^+:\Gamma(S_X^+\otimes E)\longrightarrow
 \Gamma(S_X^-\otimes E).
 \label{eq:app-dirac-even-dimension}
\end{equation}
Let a finite group \(G\) preserve the metric, spin structure, and twisting
bundle. The kernel and cokernel are then \(G\)-representations ($G$-modules), and their difference is a \textit{virtual} $G$-representation
\begin{equation}
 \operatorname{Ind}_G(D_{X,E}^+)
 =[\ker D_{X,E}^+]-[\operatorname{coker}D_{X,E}^+]
 \in R(G).
 \label{eq:app-virtual-index}
\end{equation}
Evaluation of its character at \(g\in G\) gives the $g$-equivariant index (with lower case $i$)
\begin{equation}
 \ind_g(D_{X,E}^+)
 =\Tr\!\left(g|_{\ker D_{X,E}^+}\right)
 -\Tr\!\left(g|_{\operatorname{coker}D_{X,E}^+}\right).
 \label{eq:app-equivariant-character}
\end{equation}
At \(g=1\), this character reduces to the ordinary integer-valued index in
equation~\eqref{eq:app-AS-symbol-pushforward}.

If \(Y=\partial X\) and the geometry is of product type near the boundary,
then
\begin{equation}
 D_X=\gamma^\perp(\partial_u+D_Y),
 \label{eq:app-collar-dirac}
\end{equation}
where \(D_Y\) is the self-adjoint tangential Dirac operator on the
odd-dimensional boundary. The \(4q\)-dimensional fillings appearing in parts of Donnelly's original paper
belong primarily to the signature operator. The spin--Dirac formula used
here requires only that \(X\) be even-dimensional, so that
\(S_X=S_X^+\oplus S_X^-\); it does not require
\(\dim X\in4\Z\) \cite{Donnelly,BravermanMaschler}.

We first ignore the boundary and derive the bulk fixed-point density by
equivariant \(K\)-theory. We then return to the spectral boundary term in
Donnelly's formula.

\subsection{One complex normal coordinate and the Koszul complex}
\label{app:k-theory-denominator}

We begin with the zero section of a complex line bundle \(L\to F\). Let
\(\pi:L\to F\) be the projection and let
\(v\in\Gamma(L,\pi^*L)\) be the tautological section. The \(K\)-theory Thom
class of \(L\) is represented by the Koszul complex
\begin{equation}
 0\longrightarrow\pi^*L^*
 \xrightarrow{\,\iota_v\,}\one\longrightarrow0.
 \label{eq:app-Koszul-line}
\end{equation}
Away from the zero section, \(v\ne0\) and the complex is exact. It therefore
defines a \(K\)-class supported near the zero section. Pulling it back to
the zero section sets \(v=0\), so its differential vanishes and its virtual
class becomes
\begin{equation}
 \one-L^*=\lambda_{-1}(L^*).
 \label{eq:app-line-Euler-class}
\end{equation}
Thus one complex normal coordinate contributes the \(K\)-theory Euler
factor \(1-L^*\).

For a rank-\(r\) complex vector bundle \(V\), the graded tensor product of
the one-line Koszul complexes gives
\begin{equation}
 \lambda_{-1}(V^*)
 :=\sum_{q=0}^{r}(-1)^q[\Lambda^qV^*].
 \label{eq:app-lambda-minus-one}
\end{equation}
Consequently, for the zero section \(s:F\hookrightarrow V\),
\begin{equation}
 \boxed{
 s^*s_!(a)=a\otimes\lambda_{-1}(V^*).}
 \label{eq:app-K-self-intersection}
\end{equation}
This is the self-intersection formula: after a class on the zero section is
pushed into the total space and restricted back, every complex normal
coordinate supplies one additional factor \(1-L^*\).

\subsection{The principal symbol and self-intersection on the cotangent bundle}

Specializing the general definition in
equation~\eqref{eq:app-general-principal-symbol} to the first-order twisted
chiral Dirac operator gives
\begin{equation}
 \sigma(D_{X,E}^+)(x,\xi)
 =\mathrm i\,c_x(\xi)\otimes\one_E:
 S_{X,x}^+\otimes E_x\longrightarrow S_{X,x}^-\otimes E_x.
 \label{eq:app-principal-symbol-map}
\end{equation}
The Clifford relation implies
\(c_x(\xi)^2=-\lVert\xi\rVert^2\), so this map is invertible whenever
\(\xi\ne0\). Its failure of invertibility is therefore confined to the zero
section, and the pair of pulled-back bundles together with this map defines
the compactly supported equivariant class
\begin{equation}
 [\sigma(D_{X,E}^+)]\in K_{G,c}^0(T^*X).
 \label{eq:app-symbol-class}
\end{equation}
This is the only use of the principal symbol needed below: it packages the
ellipticity of \(D_{X,E}^+\) into the \(K\)-class to which localization and
the index pushforward apply.

Fix \(g\in G\) and let \(F\subset X^g\) be a connected fixed component.
After choosing an invariant metric, there is a cotangent-bundle embedding
\begin{equation}
 \jmath_F:T^*F\hookrightarrow T^*X.
 \label{eq:app-cotangent-inclusion}
\end{equation}
If \(N_F\) is the real normal bundle of \(F\hookrightarrow X\), the real
normal bundle of \(T^*F\hookrightarrow T^*X\) contains both the displacement
normal to the base and the corresponding normal momentum:
\begin{equation}
 \nu_{\jmath_F}\simeq N_F\oplus N_F^*
 \simeq N_F\otimes\C.
 \label{eq:app-symbol-space-normal}
\end{equation}
Hence the Euler class relevant to the index proof is
\(\lambda_{-1}(N_{F,\C}^*)\). The self-intersection formula becomes
\begin{equation}
 \jmath_F^*\jmath_{F!}(a)
 =a\otimes\lambda_{-1}(N_{F,\C}^*).
 \label{eq:app-symbol-self-intersection}
\end{equation}

We next explain why no localized contribution remains away from the fixed
set. For a nonabelian finite group one first restricts to the centralizer
\(C_G(g)\); in the present \(G=\Z_k\) application this does not change the
group. Let \([g]\) denote localization of the complexified representation
ring at the maximal ideal determined by evaluation at \(g\), and put
\(U=X\setminus X^g\). An equivariant cell in \(U\) has the form
\(C_G(g)/H\) with \(g\notin H\). Since
\(K_{C_G(g)}^0(C_G(g)/H)\simeq R(H)\), localization at \([g]\) kills this
module: the character at \(g\) cannot be detected by a stabilizer that does
not contain \(g\). Applying this orbit-cell argument to the compactly
supported cotangent bundle gives
\begin{equation}
 K_{C_G(g),c}^0\!\left(T^*U\right)_{[g]}=0.
 \label{eq:app-K-away-fixed-vanishes}
\end{equation}
The exact sequence for the pair \((T^*X,T^*U)\) then says that every
localized symbol class comes from a class supported in a tubular
neighborhood of \(T^*X^g\). The Thom isomorphism identifies that supported
group with the equivariant \(K\)-theory of the fixed cotangent bundle.
Consequently, the index pushforward factors as
\begin{equation}
 K_{C_G(g),c}^0(T^*X)_{[g]}
 \longrightarrow
 \bigoplus_{F\subset X^g}K_{C_G(g),c}^0(T^*F)_{[g]}
 \xrightarrow{\ (p_F)_!\ }
 R(C_G(g))_{[g]}.
 \label{eq:app-fixed-pushforward-factorization}
\end{equation}
After applying the equivariant Chern character and Riemann--Roch theorem,
this factorization becomes an integral over the fixed components \(F\).
Thus fixed-point localization of the index density follows from the support
of the localized symbol class; it is not an additional analytic
assumption.

To see why this class can be divided out, let \(q_\alpha\) be the
eigenvalue of \(g\) on a complex normal line \(L_\alpha\). Since \(F\) is
a component of the fixed set, there is no normal direction with
\(q_\alpha=1\). At zero curvature,
\begin{equation}
 \ch_g\!\left(\lambda_{-1}(N_{F,\C}^*)\right)
 =\prod_\alpha(1-q_\alpha^{-1})\ne0.
 \label{eq:app-invertible-Euler-class}
\end{equation}
The Euler class is therefore invertible after localization of the
representation ring at \(g\). The Atiyah--Segal localization theorem gives
\begin{equation}
 \boxed{
 [\sigma(D_{X,E}^+)]_{(g)}
 =\sum_{F\subset X^g}\jmath_{F!}\!\left(
 \frac{\jmath_F^*[\sigma(D_{X,E}^+)]}
 {\lambda_{-1}(N_{F,\C}^*)}\right).}
 \label{eq:app-localized-symbol}
\end{equation}
The denominator has an elementary meaning. If
\(\jmath_F^*\alpha=b\,\lambda_{-1}(N_{F,\C}^*)\), then the class that must
actually be pushed forward from the fixed set is
\(b=\jmath_F^*\alpha/\lambda_{-1}(N_{F,\C}^*)\).

\subsection{Restriction of the Dirac symbol and the normal factor}

On a fixed component there are orthogonal decompositions
\begin{equation}
 TX|_F=TF\oplus N_F,
 \qquad
 S_X|_F\simeq S_F\mathbin{\widehat\otimes}S(N_F).
 \label{eq:app-spinor-splitting}
\end{equation}
After restriction to \(T^*F\), Clifford multiplication acts only on the
tangent spinors \(S_F\). Therefore
\begin{equation}
 \jmath_F^*[\sigma(D_{X,E}^+)]
 =[\sigma(D_{F,E|_F}^+)]\otimes
 \bigl([S^+(N_F)]-[S^-(N_F)]\bigr).
 \label{eq:app-restricted-Dirac-symbol}
\end{equation}
Combining this equation with \eqref{eq:app-localized-symbol}, the normal
\(K\)-class on the fixed set is
\begin{equation}
 \frac{[S^+(N_F)]-[S^-(N_F)]}
 {\lambda_{-1}(N_{F,\C}^*)}.
 \label{eq:app-normal-K-class}
\end{equation}
The numerator and denominator therefore have distinct origins: the
numerator comes from the chirality decomposition of the restricted Dirac
symbol, while the denominator comes from self-intersection on the cotangent
bundle.

Let \(\mathsf R_N\) be the normalized normal curvature and
\(\mathsf R_S\) its induced action on the normal-spinor bundle. Recall first
that for an equivariant bundle \(V\),
\begin{equation}
 \ch_g(V)=\Tr_V\!\left(g_Ve^{\mathsf R_V}\right).
 \label{eq:app-equivariant-ch-definition}
\end{equation}
For the normal factors below we use the dual-normal convention
inherited from \(N_{F,\C}^*\). Thus \(\mathsf R_N\) denotes the
curvature on the original normal bundle, while the exponent in the
dual Thom class is \(-\mathsf R_N\). Likewise, the symbol
\(\mathsf R_S\) in
\(\STr(\widetilde g e^{-\mathsf R_S})\) is the spinor curvature in
this dual convention. If \(\mathsf R_S^{\rm std}\) is instead defined
by the standard formula
\(\ch_{\widetilde g}(S)=\STr(\widetilde g e^{\mathsf R_S^{\rm std}})\),
then \(\mathsf R_S=-\mathsf R_S^{\rm std}\). This convention explains
the apparent sign difference between the general Chern-character
definition and the normal-spinor numerator.
On \(\Lambda^qV\), both the group action and the curvature are induced by
the exterior-power functor:
\begin{equation}
 \ch_g(\Lambda^qV)
 =\Tr_{\Lambda^qV}\!\left(
 \Lambda^q\!\left(g_Ve^{\mathsf R_V}\right)\right).
 \label{eq:app-equivariant-ch-exterior-power}
\end{equation}
Thus a scalar action \(g_V=z\,\one_V\) becomes \(z^q\) on
\(\Lambda^qV\); for a general action, the eigenvalues are products of \(q\)
distinct eigenvalues of \(g_V\). This is exactly what the
exterior-algebra identity
\begin{equation}
 \sum_q(-1)^q\Tr_{\Lambda^qV}(\Lambda^qA)=\det_V(1-A)
 \label{eq:app-exterior-identity}
\end{equation}
encodes. Applying it first on the dual normal bundle gives the unambiguous
formula
\begin{equation}
 \ch_g\!\left(\lambda_{-1}(N_{F,\C}^*)\right)
 =\det_{N_{F,\C}^*}\!\left(1-g_{N^*}e^{\mathsf R_{N^*}}\right)
 =\det_{N_F\otimes\C}\!\left(1-g\,e^{-\mathsf R_N}\right).
 \label{eq:app-normal-thom-denominator}
\end{equation}
In the last expression, the action and curvature on the dual have been
transported to \(N_F\otimes\C\) with an invariant metric; this convention is
what the symbols \(g\) and \(-\mathsf R_N\) denote there. If \(u_\alpha\)
and \(x_\alpha\) are instead defined as the eigenvalues and Chern roots on
the original normal bundle, the corresponding dual factors are
\(u_\alpha^{-1}\) and \(-x_\alpha\). The numerator is
\begin{equation}
 \ch_{\widetilde g}\!\left(S^+(N_F)-S^-(N_F)\right)
 =\STr_{S(N_F)}\!\left(\widetilde g\,e^{-\mathsf R_S}\right),
 \label{eq:app-normal-spin-numerator}
\end{equation}
where \(\widetilde g\) is the lift of \(g\) to the normal-spinor bundle.
Hence
\begin{equation}
 \boxed{
 \mathcal K_{\widetilde g}(N_F)
 =\frac{\STr_{S(N_F)}\!\left(\widetilde g\,e^{-\mathsf R_S}\right)}
 {\det_{N_F\otimes\C}\!\left(1-g\,e^{-\mathsf R_N}\right)}.}
 \label{eq:app-supertrace-over-determinant}
\end{equation}
This is the central result of the equivariant \(K\)-theory derivation:
self-intersection and restriction of the Dirac symbol already fix the ratio
of the normal denominator to the spinor numerator.

Under the splitting principle, decompose \(N_F\otimes\C\) into oriented
real two-planes. Let \(x_\alpha\) be a curvature root,
\(\theta_\alpha\) the rotation angle of \(g\), and define
\begin{equation}
 Y_\alpha=x_\alpha+\mathrm i\theta_\alpha.
 \label{eq:app-equivariant-root}
\end{equation}
With compatible orientation and spin-lift conventions,
\begin{align}
 \det_{N_F\otimes\C}\!\left(1-g e^{-\mathsf R_N}\right)
 &=(-1)^q\prod_{\alpha=1}^{q}
 4\sinh^2\!\left(\frac{Y_\alpha}{2}\right),
 \label{eq:app-split-normal-denominator}\\
 \STr_{S(N_F)}\!\left(\widetilde g e^{-\mathsf R_S}\right)
 &=\varepsilon_S(\widetilde g)\prod_{\alpha=1}^{q}
 2\sinh\!\left(\frac{Y_\alpha}{2}\right).
 \label{eq:app-split-spin-numerator}
\end{align}
It follows that
\begin{equation}
 \mathcal K_{\widetilde g}(N_F)
 =\varepsilon(\widetilde g)
 \prod_{\alpha=1}^{q}
 \frac1{2\sinh\!\left((x_\alpha+\mathrm i\theta_\alpha)/2\right)},
 \qquad \rank_{\Rset}N_F=2q,
 \label{eq:app-general-normal-factor}
\end{equation}
where \(\varepsilon(\widetilde g)=\pm1\) records the combined orientation
and spin-lift convention. The local density on the fixed component is
\begin{equation}
 \boxed{
 \mathcal I_{g,F}(D_{X,E}^+)
 =\left[\widehat A(TF)\,
 \mathcal K_{\widetilde g}(N_F)\,
 \ch_g(E|_F)\right]_{\dim F}.}
 \label{eq:app-donnelly-local-density}
\end{equation}

\subsection{The boundary spectral term in Donnelly's formula}

Equivariant \(K\)-theory has now determined the bulk fixed-point density.
For a manifold with boundary, the orbifold projector must also be inserted
in the spectral trace of the boundary Dirac operator:
\begin{equation}
 \Pi_G=\frac1{|G|}\sum_{g\in G}U_g,
 \qquad
 \eta_{Y/G}(D_Y,s)
 =\frac1{|G|}\sum_{g\in G}\eta_g(D_Y,s),
 \label{eq:app-group-projector}
\end{equation}
where
\begin{equation}
 \eta_g(D_Y,s)
 :=\Tr\!\left(U_gD_Y|D_Y|^{-s-1}\right),
 \qquad
 h_g(D_Y):=\Tr\!\left(U_g|_{\ker D_Y}\right),
 \qquad
 \xi_g:=\frac{\eta_g+h_g}{2}.
 \label{eq:app-equivariant-eta}
\end{equation}
If a flat connection is specified by \(\rho:G\to H\), \(U_g\) contains
both the geometric lift and the action of \(\rho(g)\) on the fiber. This is
why the group character remains in the equivariant Chern character.

Donnelly's equivariant APS theorem combines the bulk density with the
boundary spectral term:
\begin{equation}
 \boxed{
 \ind_g(D_{X,E}^+)
 =\sum_{F\subset X^g}\int_F
 \mathcal I_{g,F}(D_{X,E}^+)
 -\frac{\eta_g(D_{Y,E})+h_g(D_{Y,E})}{2}.}
 \label{eq:app-donnelly}
\end{equation}
Equivariant \(K\)-theory proves the form of \(\mathcal I_{g,F}\). The last
term comes from the APS spectral boundary condition and is not the Chern
character of a topological \(K\)-class. This is why equivariant
\(K\)-theory simplifies the bulk fixed-point formula but does not replace
Donnelly's boundary analysis. The exponentiated \(\eta\)-phase pairs with
the boundary Pfaffian line to form the well-defined Dai--Freed inflow system
\cite{DaiFreed,WittenYonekura,YonekuraDaiFreed}.

One group element \(g\) and one fixed component \(F\) give only
\begin{equation}
 I_{g,F}
 =\left[\widehat A(TF)\mathcal K_{\widetilde g}(N_F)
 \ch_g(E|_F)\right]_{d+2}.
 \label{eq:app-single-sector-anomaly}
\end{equation}
The orbifold projector requires the average over all group elements:
\begin{equation}
 \boxed{
 I_{G,F}
 =\frac1{|G|}
 \sum_{\substack{g\in G,\ g\ne1\\F\subset X^g}}I_{g,F},
 \qquad
 I_G^{\rm 1-loop}
 =\frac1{|G|}I_{1,X}+\sum_F I_{G,F}.}
 \label{eq:app-complete-group-sum}
\end{equation}
The identity term is \(1/|G|\) times the ordinary APS density on the cover,
while the non-identity terms are localized corrections on singular fixed
components. For \(G=\Z_k\) and the common fixed component \(F=W_6\),
\begin{equation}
 \boxed{
 I_{\Z_k,W}(E;\rho)
 =\frac1k\sum_{j=1}^{k-1}
 \left[\widehat A(TW)K_j(r)
 \ch_{g^j,\rho(g^j)}(E|_{W_6})\right]_8.}
 \label{eq:app-complete-ALE-sum}
\end{equation}
Here the map from the fixed component to six-dimensional spacetime is the
identity, so there is no additional internal pushforward.

\subsection{Angles and the normal factor on
\texorpdfstring{\(\C^2/\Z_k\)}{C2/Zk}}
\label{app:normal-angle-conventions}

For the \(A_{k-1}\) local geometry, the group action is
\begin{equation}
 g^j:(z_1,z_2)\longmapsto
 (e^{\mathrm i\theta_j}z_1,e^{-\mathrm i\theta_j}z_2),
 \qquad
 \theta_j=\frac{2\pi j}{k},
 \qquad j=1,\ldots,k-1.
 \label{eq:app-ALE-angles}
\end{equation}
The discrete rotation angle \(\theta_j\), the degree-two normal curvature
root \(r\), and the spin lift \(\widetilde g^j\) are distinct data. Choose
the standard lift compatible with the complex orientation and
\(\widetilde g^0=1\), and set
\begin{equation}
 Y_{j,\pm}=r\pm\mathrm i\theta_j.
 \label{eq:app-ALE-equivariant-roots}
\end{equation}
The numerator and denominator of
\eqref{eq:app-supertrace-over-determinant} are
\begin{align}
 \mathcal N_j(r)
 &=\STr_{S(N)}\!\left(\widetilde g^j e^{-\mathsf R_S}\right)
 =4\sinh\!\left(\frac{Y_{j,+}}2\right)
   \sinh\!\left(\frac{Y_{j,-}}2\right),
 \label{eq:app-ALE-spin-numerator}\\
 \mathcal D_j(r)
 &=\det_{N\otimes\C}\!\left(1-g^j e^{-\mathsf R_N}\right)
 =16\sinh^2\!\left(\frac{Y_{j,+}}2\right)
    \sinh^2\!\left(\frac{Y_{j,-}}2\right).
 \label{eq:app-ALE-vector-denominator}
\end{align}
Their ratio is
\begin{equation}
 K_j(r)=\frac{\mathcal N_j(r)}{\mathcal D_j(r)}
 =\frac1{
 4\sinh\!\left(\frac{r+\mathrm i\theta_j}{2}\right)
  \sinh\!\left(\frac{r-\mathrm i\theta_j}{2}\right)}
 =\frac1{2(\cosh r-\cos\theta_j)}.
 \label{eq:app-ALE-normal-factor}
\end{equation}
In particular,
\begin{equation}
 K_j(0)=\frac1{4\sin^2(\theta_j/2)},
 \qquad
 K_{k-j}(r)=K_j(r).
 \label{eq:app-ALE-flat-limit}
\end{equation}
Thus \(K_j(r)\) is obtained by multiplying the \(1-L^*\) factors of the
complex normal coordinates and dividing the resulting self-intersection
class into the normal-spinor virtual class.

This appendix uses equivariant \(K\)-theory localization for the spin--Dirac
symbol. For the signature complex, Scrucca--Serone gave a
physics-oriented derivation of the equivariant signature theorem, which is
complementary to the spin--Dirac symbol argument used here
\cite{ScruccaSerone1999}.

\section*{Declaration on the use of artificial intelligence}
Generative artificial intelligence tools were extensively used in the
analysis and writing of this article. The author has verified all results
and takes full responsibility for all statements made here.

\bibliographystyle{alpha}
\bibliography{references}

\end{document}